\documentclass[twocolumn,3p,times,procedia]{elsarticle}

\usepackage{ecrc}
\usepackage{graphicx}
\usepackage{titlesec}
\usepackage{array,threeparttable}
\usepackage{url}

\usepackage{multirow}
\usepackage{booktabs}
\usepackage{makecell}
\usepackage{hyperref}
\usepackage[hyphenbreaks]{breakurl}

\volume{00}

\firstpage{1}

\runauth{}

\jid{procs}

\CopyrightLine{2022}{Published by Elsevier Ltd.}

\usepackage{amssymb}
\usepackage{amsmath}
\usepackage{multicol}

\usepackage[figuresright]{rotating}
\usepackage{bm}
\usepackage{caption}
\usepackage{fancyhdr}

\fancyheadoffset[RO,EL]{0pt}
\usepackage{geometry}

\begin{document}

\begin{frontmatter}




\dochead{}
\title{
\begin{flushleft}
{\bf \Huge A new technology perspective of the Metaverse: its essence, framework and challenges}
\end{flushleft}
}
 %

\author[]{\bf \Large \leftline {Feifei Shi$^a$, Huansheng Ning$^*$$^a$$^b$, Xiaohong Zhang$^b$, Rongyang Li$^a$, Qiaohui Tian$^a$,}\bf \leftline {Shiming Zhang$^a$, Yuanyuan Zheng$^a$, Yudong Guo$^b$, Mahmoud Daneshmand$^c$}}



\address{\bf  \leftline {$^a$School of Computer and Communication Engineering, University of Science and Technology Beijing, Beijing, 100083, China.}

\bf  \leftline {$^b$Jinzhong University, Shanxi, 030619, China.}

\bf  \leftline {$^c$Department of Business Intelligence
and Analytics and the Department of Computer Science, Stevens Institute of Technology,}
\bf  \leftline{Hoboken, NJ 07030, USA.}
}

\cortext[]{Corresponding author: Huansheng Ning (ninghuansheng@ustb.edu.cn)}



\begin{abstract}

The Metaverse depicts a parallel digitalized world where virtuality and reality are fused. It has economic and social systems like those in the real world and provides intelligent services and applications. In this paper, we introduce the Metaverse from a new technology perspective, including its essence, corresponding technical framework, and potential technical challenges. Specifically, we analyze the essence of the Metaverse from its etymology and point out breakthroughs promising to be made in time, space, and contents of the Metaverse by citing Maslow's Hierarchy of Needs. Subsequently, we conclude four pillars of the Metaverse, named ubiquitous connections, space convergence, virtuality and reality interaction, and human-centered communication, and establish a corresponding technical framework. Additionally, we envision open issues and challenges of the Metaverse in the technical aspect. The work proposes a new technology perspective of the Metaverse and will provide further guidance for its technology development in the future.

\end{abstract}


\end{frontmatter}



\section{Introduction}

To the best of our knowledge, the continuous development of the Internet of Things (IoT) and cyberspace has established ubiquitous connections between objects, things, and humans. Meantime, it also promotes the seamless convergence among physical, social, thinking, and cyber spaces \cite{A1}. In recent years, such human activities as work, shopping, conference, and entertainment have been increasingly changed into online forms. Particularly in the context of the COVID-19 pandemic, people tend to spend more time in virtual space. New commercial forms enable more industries to seek innovative developing ways, especially in these pioneering realms like electronic games, fashions, education, etc. 

The Metaverse thrives at this stage with the vision to provide more possibilities in daily life and industrial manufacturing \cite{A2}. It has become such a hot buzzword these days and has attracted lots of attention from both academia and the industry. We are all impressed by the wide spread of Metaverse, from the first science fiction \emph{Snowcrash} to Facebook, the giant IT company that was even renamed Meta a few months ago. Compared with the Internet world or cyberspace which refers to a network of networks, the Metaverse depicts a parallel and immersive world where virtuality and reality are fused. It could be regarded as a hypothesized iteration of cyberspace, where humans could enter the virtual world with techniques such as Virtual Reality (VR), Augmented Reality (AR), etc \cite{A3}. The vision of the Metaverse is to provide an immersive user experience with low latency and strong intelligence. We can imagine that the scenes in the film of \emph{Ready Player One} where everyone can be interconnected with the world of OASIS, will come true in Metaverse \cite{A4}. However, the digitalized Metaverse is not only for playing games, but also portrayed as a digital world with persistence and synchronization, as well as the economy, culture, regulation, ethics, and morality. Hence, to support complicated applications, the Metaverse must be equipped with advanced techniques to keep activities, interactions, and transactions safe, transparent, and sustainable \cite{A5}.

\begin{figure*}[!ht]
\centering
\includegraphics[width=17cm]{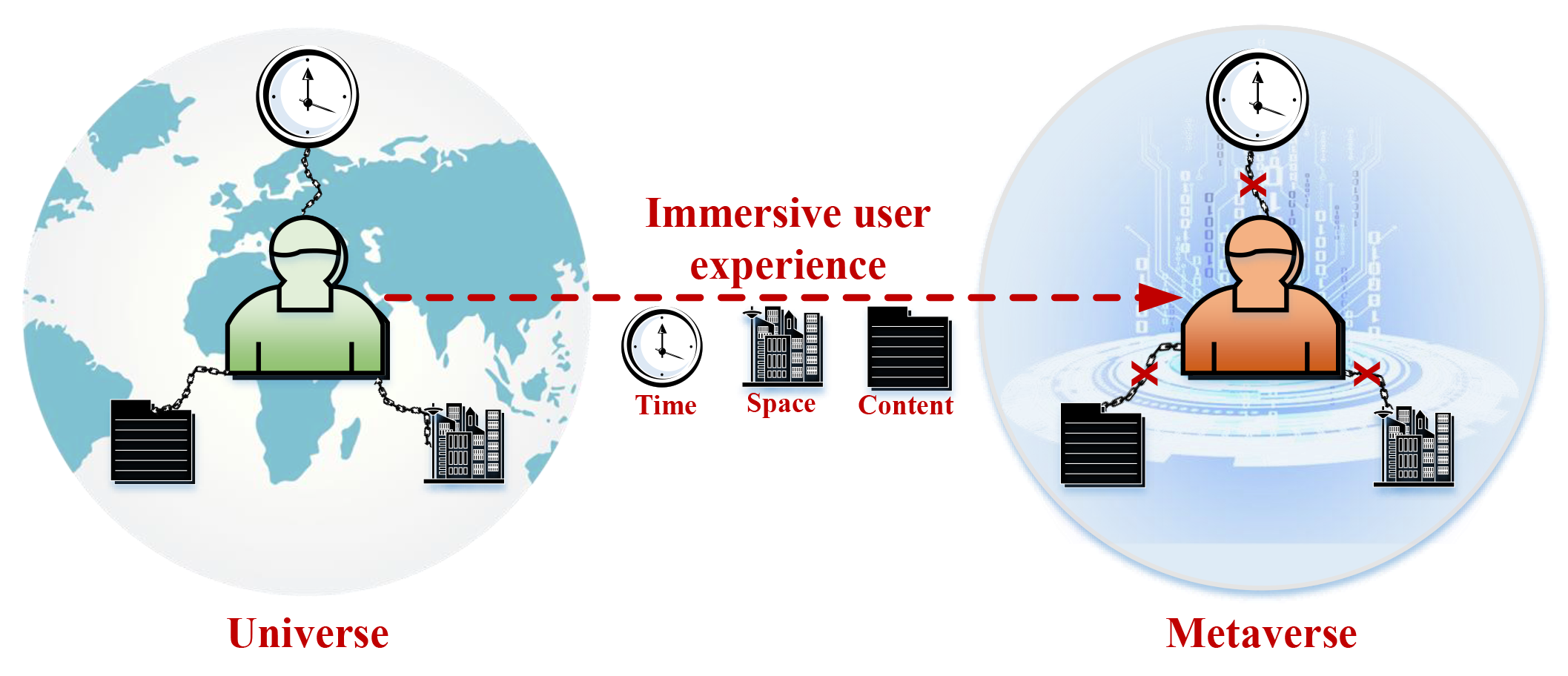}
\caption{The differences between the essence of the universe and the Metaverse.}
\label{2}
\end{figure*}

To create a deeply virtualized Metaverse and fulfill the extremely immersive user experience, it is necessary to carry out innovative technical research. Many scholars and engineers have launched exploratory research, particularly research on supporting techniques in the Metaverse. For example, in \cite{A6}, the authors mention that the Metaverse is not a composition of one or more technologies. It depends on six technical pillars named BIGANT, that is Blockchain, Interactivity, Game, Artificial Intelligence (AI), Network, and IoT. Lee also overviews the technological singularity of Metaverse from the aspects of Extended Reality, User Interactivity, AI, Blockchain, Computer Vision, IoT and Robotics, Edge and Cloud computing, and Future Mobile Networks \cite{A7}. All these techniques will contribute a lot to Metaverse development, while most of them are proposed according to specific application requirements.


Different from existing studies, this article introduces the Metaverse from a new technical perspective, ranging from its essence, and technical framework to open challenges. The main contributions of this paper are as follows:

\begin{itemize}
\item Introducing the essence of Metaverse from its etymology, and illustrating the fact that Metaverse needs to go beyond the limits of space, time, and contents with the case of Maslow's Hierarchy of Needs.
\item Proposing four pillars of the Metaverse named ubiquitous connections, space convergence, virtuality and reality interaction, and human-centered communication, and establishing a corresponding technical framework.
\item Concluding further technical challenges in the Metaverse with guidance for upcoming research.
\end{itemize}

The remainder of this paper is arranged as followed: Section 2 introduces the essence of the Metaverse from its etymology. Section 3 proposes the four pillars of the Metaverse and establishes the corresponding technical framework. Section 4 points out some open issues related to technical development in the Metaverse. Section 5 concludes this paper.

\begin{table*}[!ht]
\centering
\caption{Some wonderful experiences related to Maslow's different hierarchies of needs in the Metaverse.}
\label{tab1}
\begin{tabular}{lllll}
\toprule
\toprule
\makecell[l]{Hierarchy \\ of Needs}  &  Case  &  Space  &   Time   &   Content  \\
\midrule
\makecell[l]{Physiological \\ Needs} & Food &  \multicolumn{1}{m{4cm}}{In the Metaverse, it would be easy to acquire precious food ingredients all over the world. Their cultivation will not be severely limited by geographical location, temperature, humidity, etc.}   & \multicolumn{1}{m{3cm}}{The food in the Metaverse is rarely subject to strict shelf-life restrictions, and some may be kept fresh forever.}   &  \multicolumn{1}{m{3cm}}{In the metaverse, food does not necessarily need to be processed or made from ingredients. Instead, modern technologies are adopted to simulate the smells, tastes and colors of food.}  \\
\midrule
Safety needs & Transportation   &   \multicolumn{1}{m{4cm}}{It allows human to travel around the world even if they are at home. Users could break the speed limit and realize instantaneous movement at their will. Additionally, it is possible to travel to places with harsh environments, such as Antarctica and Arctic, icebergs and volcanoes.}   &   \multicolumn{1}{m{3cm}}{One of the key features in the Metaverse is time travel which can provide wonderful user experience. People can go back to the past or fly to the future in the Metaverse.}   &  \multicolumn{1}{m{3cm}}{The transportation in the Metaverse may not only depend on vehicles, ships, and airplanes, but also some advanced ways through brainwaves, specific actions, etc.} \\
\midrule
\makecell[l]{Love and \\ belongingness \\ need}    &   Friendship   &   \multicolumn{1}{m{4cm}}{The humans could establish friendship and communicate with each other face to face in the Metaverse, even if the two sides are thousands of miles apart in the real world.}   &   \multicolumn{1}{m{3cm}}{The friendship in the Metaverse will not depend too much on time constraints. For example, the impact of time difference on social interaction is so tiny that could be neglected.}   &  \multicolumn{1}{m{3cm}}{The types of social friendship are various in the Metaverse. It could be established between humans and virtual identities, and also could be among virtual identities.} \\
\midrule
Esteem needs    &   \makecell[l]{Social \\ position}   &   \multicolumn{1}{m{4cm}}{Humans could hold social positions in the Metaverse similar to that in the real world. In addition, human beings can have more social status in different countries, fields and spaces with competent capabilities.}   &   \multicolumn{1}{m{3cm}}{Humans can play different social roles in the Metaverse, no matter the roles in the ancient times or in the future. For example, one can travel back to the ancient times to play Chinese emperors.}   &  \multicolumn{1}{m{3cm}}{There may be various social organizations in the Metaverse, which may be completely virtual or real. Humans could hold different positions, acquire respect, and satisfy self-esteem needs in the Metaverse.} \\
\midrule
\multicolumn{1}{m{2cm}}{Self-actualization needs}    &   \makecell[l]{Self-\\fulfillment}   &   \multicolumn{1}{m{4cm}}{The Metaverse provides more opportunities for humans to achieve self-fulfillment. For example, by finding friends with similar talents in the Metaverse, humans could receive recognition for their specific skills and realize self-fulfillment.}   &   \multicolumn{1}{m{3cm}}{The Metaverse may have detailed records for any achievements in the past, and keep the feeling of self-fulfillment away from fading over time.}   &  \multicolumn{1}{m{3cm}}{In the Metaverse, humans have multiple ways to achieve self-fulfillment. They are able to create cultural, psychological, and other virtual contents and values in the Metaverse.} \\
\bottomrule
\bottomrule
\end{tabular}
\end{table*}

\section{The Essence of Metaverse from its Etymology}

So far, there is no precise definition of the essence of the Metaverse. Some regard it as a novel living space for humans, while others consider it a combination of multiple technologies. Although people remain divided over the essence of the Metaverse, they have reached a consensus that Metaverse offers more possibilities for immersive user experiences. In other words, the Metaverse changes the way we interact with virtual environments. It claims to improve the immersion largely with advanced techniques. 

The word ``Metaverse'' is composed of the prefix ``Meta'' and the suffix ``verse''. As we all know, the ``Meta'' was a Greek term, and it is popularly used as a prefix to mean after or beyond. For example, the word ``metadata'' often means something more than data, especially with a self-referential connotation. The term ``verse'' is the abbreviation of ``universe''. Inspired by it, we argue that the portmanteau word ``Metaverse'' could be regarded as a computer-generated virtual space that is beyond the universe \cite{A8}. If it is, the essence of the Metaverse should also go beyond that of the universe.

According to the physical or philosophical definition, the essence of the universe could be understood as all of space, time, and their contents, including various matter and energy \cite{A117}. We can say that user experiences in the universe or the real world are restricted by spatial-temporal coordinates. For example, any true feeling in daily life must occur at a given time and place. Additionally, in most cases, the experience in the real world depends on actual contents, such as the astonishing culture shock produced when one watches an exhibition in a museum, the rumblings of workplaces enabling one to realize the process of industrial manufacturing, etc. Moreover, The user experience must abide by some rules in real life, e.g., time past cannot be called back; physical speed has upper limitations.


Based on the etymology of the Metaverse, we infer that the essence of the Metaverse goes beyond that of the universe, especially in the aspects of space and time. Figure \ref{2} depicts the differences between the essence of the universe and the Metaverse. To provide immersive user experiences, the Metaverse focuses on breaking the limits of space, time, and content. It changes the way humans interact with the outside and concentrates more on the enhancement of immersion. 

In other words, there would be less dependence on spatial and temporal characteristics, and all the contents are much more abundant and feasible. For example, with the help of holographic projection, the late Teresa Teng could ``stand'' on the stage and sing with other singers at the concert, which provides a visual feast with hyper spatiotemporality. Such scenes without physical and spatial restrictions would be very common in the Metaverse. To provide a clear illustration of the changes in the future Metaverse, we cite Maslow's Hierarchy of Needs and describe the possible progress that the Metaverse brings in respective stages of needs.

To the best of our knowledge, the popular five-stage Maslow's Hierarchy of Needs includes physiological needs, safety needs, love and belongingness needs, esteem needs, and self-actualization needs \cite{A119}. Physiological needs refer to the biological requirements for human survival, such as food, shelter, and clothing. Once these basic needs are satisfied, safety needs become salient. These needs include the security of the body, employment, resources, and health. The love and belongingness needs are the third level in the hierarchical theory, which concentrates on social feelings, especially the emotional needs of relationships. Esteem needs refer to the need for respect, self-esteem, and self-confidence. It includes not only respect for yourself, but also the desire and need for respect from others. The highest level in Maslow's Hierarchy is the self-actualization needs, which motivate humans to realize their potential and seek personal growth and self-realization. Maslow's Hierarchy of Needs describes the five most basic and innate needs, which are the motivations guiding individual behavior.

The Metaverse will provide wonderful experiences when humans strive to satisfy those requirements, especially by breaking spatial and temporal limits. According to Maslow's Hierarchy of Needs, Table \ref{tab1} lists some cases of progress in the Metaverse. For example, to meet physiological needs, advanced technology will be used to emulate the tastes and smells of food so that it is possible for people to ``experience'' virtual food in the Metaverse.

Although we introduce some cases in the Metaverse without spatial and temporal constraints, it is worth noting that hyper spatiotemporality is used to describe the features of user experience in the Metaverse. Considering that people are still in the traditional physical world, they cannot completely escape time and space constraints. In other words, we argue that the Metaverse attempts to overcome the spatial-temporal restrictions for immersive user experience. The Metaverse comes from the universe and transcends the universe, especially in terms of time and space. The blueprint of the Metaverse depicts a more immersive living space parallel to the real world for humans.

\section{Pillars in the Metaverse and the corresponding technical framework}

\begin{figure}[!ht]
\centering
\includegraphics[width=7cm]{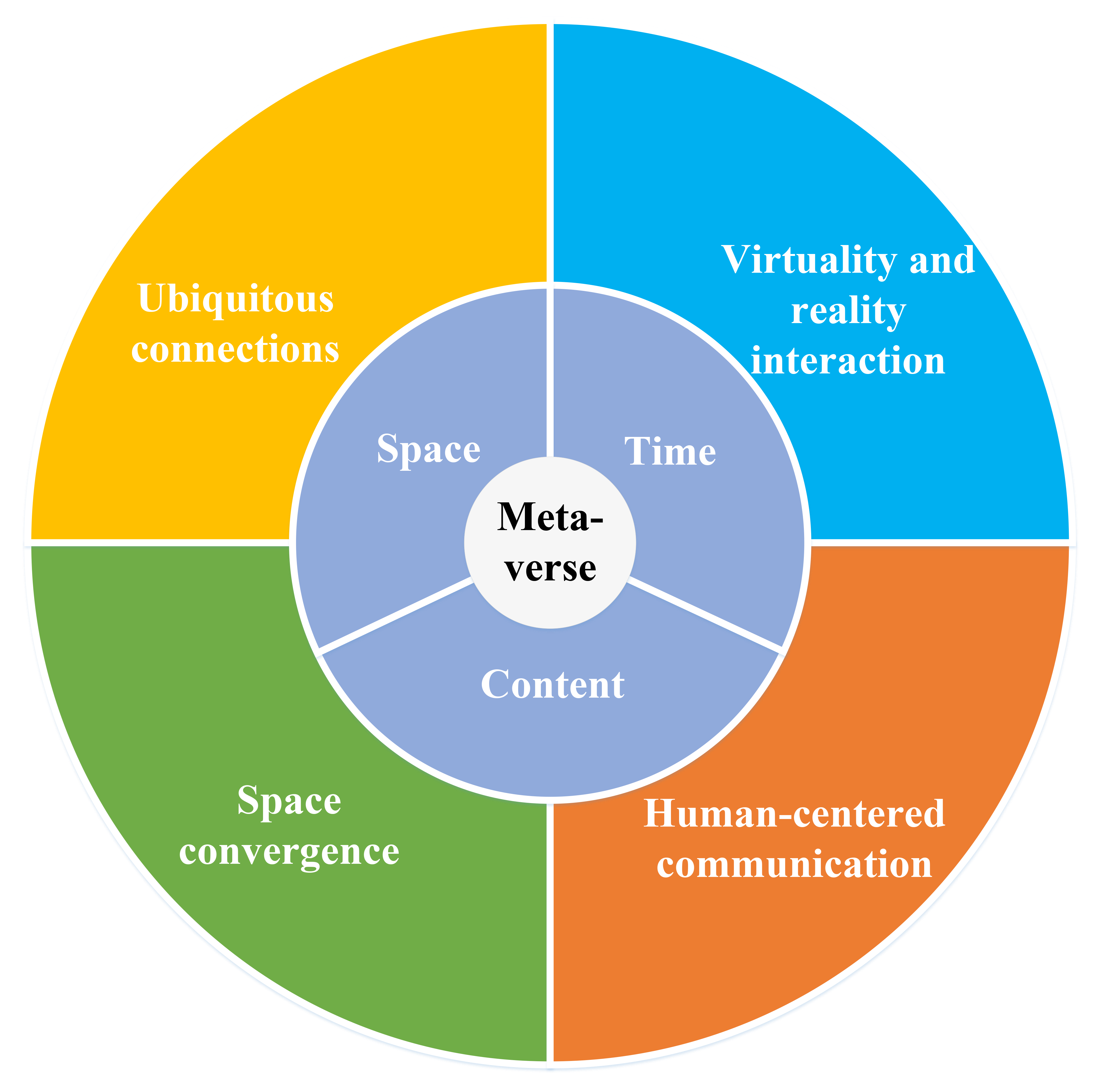}
\caption{The four pillars in the Metaverse.}
\label{fig_2}
\end{figure}

As mentioned above, the Metaverse aims at providing an immersive user experience, which needs to overcome limitations of space and time and expand contents as much as possible. Hence, advanced technologies play fundamental roles in the Metaverse. As shown in Figure \ref{fig_2}, we conclude four pillars for the Metaverse, namely ubiquitous connections, space convergence, virtuality and reality interaction, and human-centered communication. These pillars make it possible to break physical boundaries and temporal limitations and achieve an immersive user experience in the Metaverse. In this section, we would like to introduce the pillars of the Metaverse and the corresponding technical framework.

\subsection{Four pillars in the Metaverse}

In this section, we give an overall introduction to the four pillars, which largely contribute to the great immersion in the Metaverse.

\subsubsection{Ubiquitous connections}

As Chris Wysopal emphasized in Forbes, ubiquitous connection, or ubiquitous connectivity describes the fact that the connections between devices and software are so omnipresent and already exist in every corner of our life \cite{A14}. In the last years, it has witnessed the significant progress brought by the Internet, to which almost everything could be connected safely and seamlessly \cite{A9}. By establishing ubiquitous connections among things, humans, and objects in both real and virtual spaces, various activities could be carried out. For example, people can communicate without barriers even if they are thousands of miles apart; trades can be conducted in an orderly manner regardless of different time zones.

The ubiquitous connections make it possible to break spatial-temporal restrictions and lay foundations for the emergence of the Metaverse. For example, techniques related to sensing and perception provide possibilities for entities in the real world to enter the virtual world. Techniques in networking and communication allow data to transmit smoothly between different parties, overcoming the limitations in space, time, and content. Moreover, applications and services in daily life such as smart cities, intelligent transportation, and smart healthcare would be also replicated and achieved in the Metaverse. Additionally, there are more advanced applications in the Metaverse, for instance, immersive commerce, education, traveling, and so forth.



There already exist some explorations of Metaverse with the help of ubiquitous connections. For instance, Schaf establishes a prototype named 3D AutoSysLab for distant education scenarios. It designs an initial framework of an immersive learning environment with techniques like wearable sensors, mixed reality, 3D social Metaverse, and simulation modeling \cite{A10}. Han points out that sensors and devices in IoT could help replicate things in the real world to the Metaverse. This virtual replication is named digital twin, which helps a lot in dynamic resource allocation and risk prediction \cite{A11}. The digital twin also plays a significant role in smart cities and manufacturing, where the smart city digital twin is a simulation model of the physical assets, buildings, roads, and other entities in cities \cite{A12}. By replicating the real scenarios in smart cities, it is possible to monitor the present conditions, and make the appropriate response in case of any unpredictable situations \cite{A13}.

The developments of ubiquitous connections, in particular, the technical advances in IoT realize the interconnections between real and virtual space. It overcomes the limitations between spaces and provides possibilities for the Metaverse to come true and develop towards further prosperity.

\subsubsection{Space convergence}

Due to the continuous development of ubiquitous connections, an overwhelming convergence between physical, social, thinking, and cyber spaces emerges, which could be interpreted as General Cyberspace \cite{A1}. It interprets the omnipresent convergence between the Cyber-Physical system (CPS), Cyber-Physical-Social system (CPSS), and Cyber-Physical-Social-Thinking hyperspace (CPST), which is sure to break the spatial boundaries, and create possibilities for the interactions between virtual and real spaces \cite{A15,A16,A17}. In other words, the techniques of space convergence will contribute a lot to the development of the Metaverse.

Since physical space is the basic premise of all the other existence, it is the first to achieve the convergence between physical and cyber spaces. At the initial stage with simple printers, cameras, and servers, the convergence is merely realized by mutual mapping. For instance, Khanna proposes an IoT-based system for smart parking in which the practical parking resources and traffic volumes have been mapped in the cloud server to achieve real-time computing and provide appropriate suggestions \cite{A18}. In the following days, the technical advances in ubiquitous computing and ambient environment will make the convergence between physical and cyber spaces closer. For example, Tao illustrates the techniques of digital twins towards Industry 4.0 which could achieve high-fidelity cyber-physical integration with accurate models, throughout the whole process of smart manufacturing \cite{A19}. These techniques will contribute a lot to building vivid scenes in the Metaverse, and providing fundamental support for the immersive and digitalized world.

Another typical aspect of space convergence should be that between social and cyber spaces. Although the initial communication between humans is limited by physical boundaries and time constraints, the developments of the Internet have made communication more flexible. A surprisingly high number of social media such as Twitter, WeChat, LinkedIn, Instagram, Facebook, etc. have emerged whereby online communication could proceed smoothly \cite{A20}. Users on social networking sites have corresponding identities representing themselves. The virtual identities, the so-called ``avatars'' in the virtual space integrate social and cyber spaces and establish online social relationships as required. This could be regarded as the initial prototype of the Metaverse, as Mark Zuckerberg once emphasized that online social sites would strive for an extremely interconnected Metaverse in the future. The continuously technical evolution in cyber-social space convergence would be sure to promote the development of the Metaverse.

In addition, the convergence between thinking and cyber spaces also becomes popular these days, especially with the advances in disciplines of neuroscience, cognitive computing, and brain informatics. Although there still exists disputes in connecting thoughts and ideas together via the Internet, a big step forward has been made in this regard with the help of embedded sensors, electrodes, etc. They could monitor and analyze the brain signals, and then make appropriate responses with output devices. For example, in 2016, Elon Musk co-founded a company named Neuralink to develop ultra-high bandwidth Brain-computer Interfaces (BCI) \cite{A22}. In the following years, the company has built an integrated brain-machine interface platform with thousands of channels, which shows high scalability in clinical packages \cite{A21}. And it also announced a prototype of BCI named LINK V0.9 in 2019 which was implanted into the pig's brain and demonstrated high performance in monitoring the pig's brain activities \cite{A23}. These technologies not only provide the basis for the further understanding of human thoughts but also lay foundations for Metaverse development, where all ideas, thoughts, and brain activities should be understood in the digitalized space.

\subsubsection{Virtuality and reality interaction}

As the Metaverse describes a digitalized space while human bodies are still in the real physical space, the interactions between virtual and real space are extremely significant. Humans need to enter the Metaverse and enjoy wonderful experiences, in which the techniques related to virtuality and reality interaction provide appropriate ways.

Firstly, the continuous developments of interaction technologies provide the possibility to enter the Metaverse. With the iterative update of such interactive devices as mice, keyboards, touch screens, and modern wearable equipment (helmets and 3D glasses), interactive technologies have made a great step forward. For example, in \cite{A25} the authors design a 3D educational game based on VR Gear and Samsung VR display devices. Establishing online teaching and learning scenarios allows users to acquire wonderful learning experiences at home. Epp concludes common VR headsets are widely adopted in VR games, such as Oculus Rift, HTC Vive, and Windows Mixed Reality. They could help enter the virtual games more easily and improve the quality of VR games to a large extent \cite{A24}. Considering the advances in interactive technologies, it would be much more convenient for users to enter the Metaverse and enjoy fantastic services in the coming days. 

Besides, techniques related to virtuality and reality interaction create and optimize various forms of content, which may contribute a lot to immersive user experiences in the Metaverse. With the help of VR, AR, and other interactive technologies, richer contents could be extended without considering spatial-temporal constraints. For example, techniques of holographic projection contribute a lot to a digital art exhibition and show a high performance in reproducing 3D images of artworks \cite{A26}. It enables users to enjoy visual feasts by replicating and creating different contents from ancient times to the near future, from polar oceans to the ends of the universe. Especially in recent years, techniques like precise location trackers, intelligent gloves, and motion capture systems make it possible to have a much more immersive user experience, opening the way to the prosperity of Metaverse.

\subsubsection{Human-centered communication}

\begin{figure*}[!ht]
\centering
\includegraphics[width=16cm]{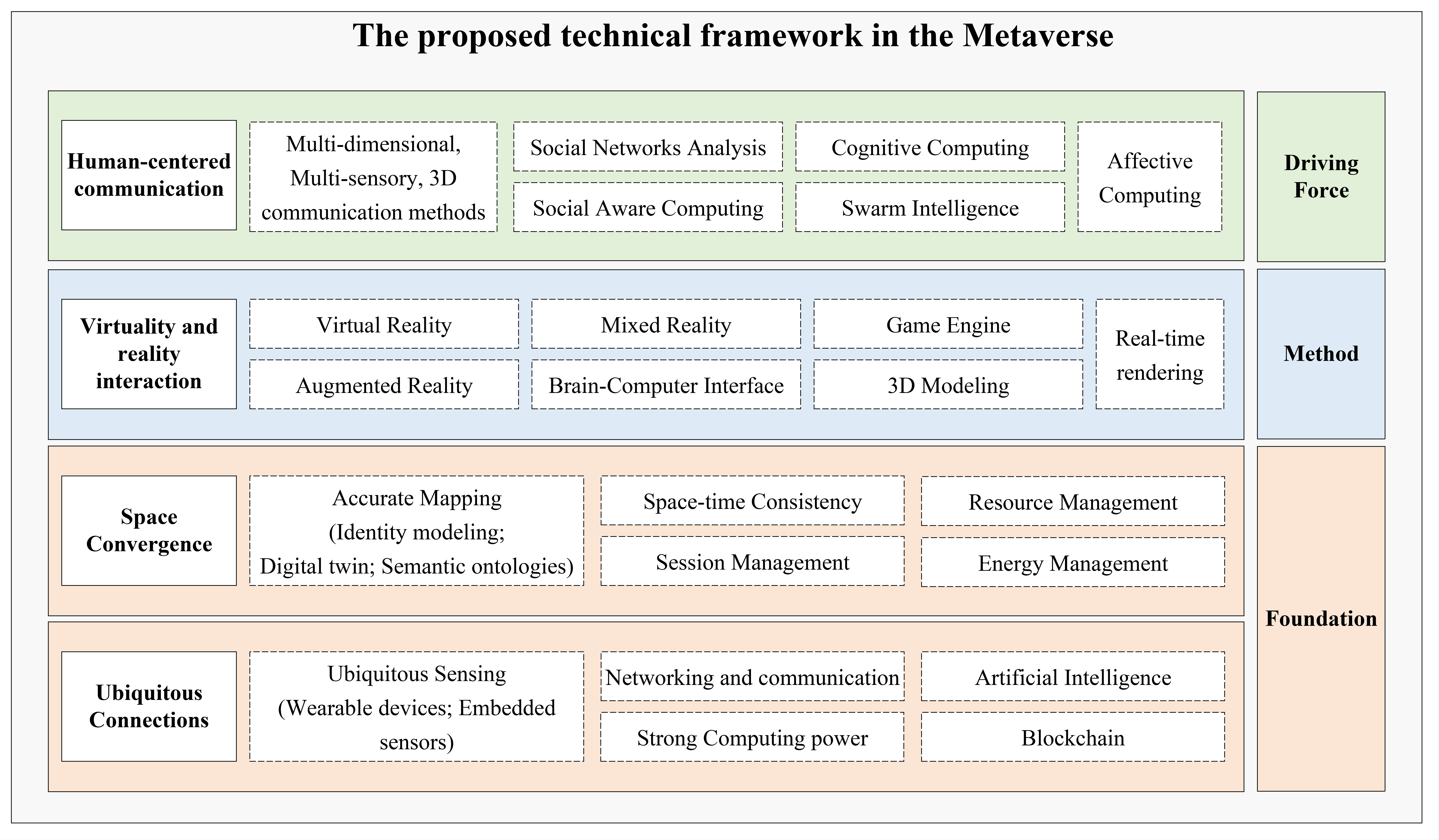}
\caption{The technical framework corresponding to the four pillars in the Metaverse.}
\label{fig_3}
\end{figure*}

Human-centered communication has been developed from the initial face-to-face communication, written letters, to wired and wireless communication, and today's instant messaging. The techniques of instant messaging have provided great convenience in humans' daily life. It allows the communication to break the spatial and temporal constraints and get a timely response. The advances in instant messaging make human-centered communication no longer limited to real space but shift gradually to virtual cyberspace. However, a case in point is that most online social communication still relies on media such as electric screens. The two sides of instant communication keep in touch through text, voice, and video, but they could not get a more immersive experience.

In the Metaverse, human-centered communication also plays a central part since most activities are carried out based on humans' communication and collaboration. To improve the quality of human-centered communication, technical breakthroughs need to be explored to further break the constraints of time and space and enhance the sense of immersion. For instance, authors in \cite{A27} adopt 3D simulation technologies and establish a system named ``Avatar to Person'', which is designed to repair virtual faces and generate a simulated voice for the disabled. It finally allows humans to communicate effectively and improves the social participation of people with disabilities. Chang makes a case study of K-Live in Korea and adopts hologram technologies to monitor user experience to achieve sustainable satisfaction and immersion \cite{A28}. This situation also appeared at Jiangsu Satellite TV New Year's Eve concert 2022, where the late Teresa Teng appeared on the stage singing and performing with other singers.

The continuous advances in human-centered communication have provided possibilities for the development of the Metaverse, and in turn, the Metaverse also puts forward technical requirements to a certain extent. When communicating with each other in the Metaverse, humans may depend on much more diverse and convenient media, or there may be no media at all. More efficient communication modes, faster transmission speed, and strong computing abilities are needed in the Metaverse for people to indulge more in the parallel digitalized world.

\subsection{The technical framework corresponding to the four pillars in the Metaverse}

Figure \ref{fig_3} depicts the proposed technical framework in the Metaverse corresponding to the four pillars. It can be seen that the techniques of ubiquitous connections and space convergence are solid foundations for overcoming the spatial-temporal constraints, and the techniques related to virtuality and reality interaction serve as methods of helping humans access the Metaverse and communicate with each other. In addition, techniques of virtuality and reality interaction could also enrich the immersive contents and scenarios in the Metaverse. Moreover, techniques related to human-centered communication are like the driving force in the Metaverse. Since humans are still at the center of the Metaverse, they put forward higher requirements for multi-dimensional, full-sensory, and 3D communication experiences. In this section, we give a comprehensive overview of the techniques involved in the framework.

\subsubsection{Potential techniques in aspect of ubiquitous connections}
\begin{figure}[!ht]
\centering
\includegraphics[width=7.5cm]{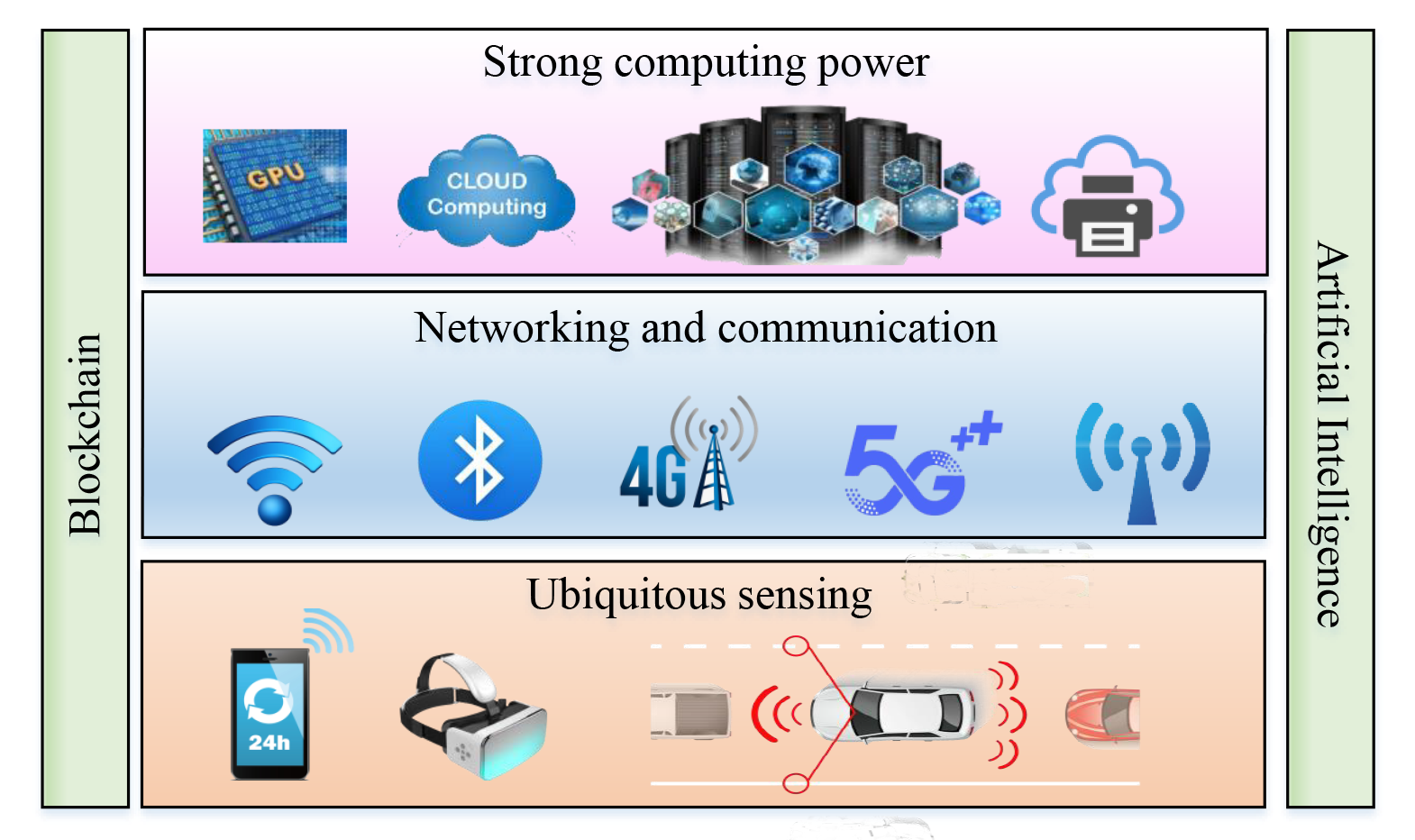}
\caption{Techniques related to ubiquitous connections.}
\label{3}
\end{figure}

As discussed above, the Metaverse is such a parallel and digitalized world that allows humans to interact with each other. Techniques of ubiquitous connections are so significant for things, humans, and entities to establish relationships. Figure \ref{3} clarifies the potential techniques related to ubiquitous sensing, which would contribute a lot to the development of the Metaverse.

\ \par
\noindent
\textbf{a) Ubiquitous sensing}
\ \par
Ubiquitous sensing refers to the abilities of omnipresent perception via different sensors attached or embedded in surroundings \cite{A89}. Similar to the sensing layer in IoT architecture, the ubiquitous sensing in the Metaverse also serves as the foundation, responsible for sensing objects and collecting information from the surroundings.

To the best of our knowledge, sensing techniques in real life mainly include sensor networks, Radio Frequency Identification, GPS positioning, etc. They rely on various sensors to acquire abundant information and further make instant decisions. For example, Li adopts ambient sensors to monitor inhabitants' daily activities, especially those of people with disabilities and the elderly in empty nests who may encounter potential risks \cite{A90}. In global positioning systems, sensing techniques allow the precise perception of location, velocity, and direction via different sensors, which have been widely used in areas of smart traffic, team sports, and logistics. 

The ubiquitous sensing is also significant in the Metaverse. It serves to collect abundant information from surroundings as required and serves as the foundation for providing immersive user experiences in the Metaverse. In addition to sensing the virtual environments, the Metaverse also needs to strengthen the control over ``avatars'', and monitor users' immersion timely. Hence, techniques of ubiquitous connections in the Metaverse are stronger in context awareness and seamless perception. There already have advanced techniques equipped with wearable devices, interactive helmets, and glasses, as well as embedded chips and electrodes. For example, the Metaverse prototype of a university campus proposed in \cite{A5} is designed to provide ubiquitous sensing-based services. It adopts location information and eye-tracking as sources of sensing input and owns the ability of ubiquitous sensing to perceive the surroundings efficiently.

The ubiquitous sensing provides access to data and information in the Metaverse and lays foundations for further applications and services. We could not imagine what would happen if the status of the ``avatar'' is not perceived accurately and timely. VR games can’t provide the best user experience without immediate and precise sensing.
\ \par
\noindent
\textbf{b) Networking and communication}
\ \par
Techniques of networking and communication in the Metaverse are responsible for safe and error-free data transmission, which is similar to the role of the network layer in IoT architecture. The most common techniques in the network layer include short-range wireless communication, and mobile communication technologies, which are also applicable in the Metaverse for networking and communication. 

As for the techniques related to short-range wireless communication, they refer to the transmission paradigm where the communication sender and receiver transmit information through radio waves, and the transmission distance is limited to a short range. Bluetooth and WiFi are the most popular representatives. As a global specification used for wireless data and voice communication, Bluetooth technology serves as special, short-distance connectivity without wires, aiming to establish a communication environment for fixed and mobile devices. It supports the frequency band of 2.45GHz used by people around the world, providing a transmission rate of 1Mbps and a transmission distance of 10m \cite{A91}. WiFi is a wireless extension of Ethernet, which is mainly used in home wireless networks and public hot spots such as airports, hotels, shopping malls, etc., while its transmission distance is easily blocked by walls \cite{A92}. These short-range wireless communications are equally significant in the Metaverse and will be adopted between devices and ``avatars''. For instance, Bluetooth would be popular in VR devices due to its small size and low cost.

Additionally, techniques of mobile communication also play significant roles in the Metaverse. Due to its fast evolution in recent years, mobile communications technology has developed from the first-generation technology to the fifth (5G) and sixth generation (6G) communication, which have brought huge benefits to human life. For example, the fourth generation (4G) communication technology mainly adopts orthogonal frequency division multiplexing and multiple-input multiple-output technologies, and it could support faster communication speed and a wider network spectrum \cite{A93}. The 5G is an extension of previous generations of technology and is currently the latest cellular mobile communication technology. Its goal is to achieve high-speed data transmission with low latency, low cost, large system capacity, and large-scale device connection \cite{A94}. With the blessing of emerging technologies, 6G emerges as a new communication technology. It is revolutionizing applications in various domains and providing immense impacts on citizens, consumers, and businesses for a future society of fully intelligent and autonomous systems \cite{A95}. Since the Metaverse is aimed at providing immersive user experiences and puts forward higher requirements in data transmission and processing, the 6G, as well as other future communication techniques, could provide intelligent capabilities of high-speed data transmission, enabling users to achieve an immersive experience to a large extent.

\ \par \noindent \textbf{c) Strong computing power}
\ \par
Frankly speaking, the Metaverse provides much more complicated applications such as high-precision VR games, immersive social sites, and complex economic systems, which require stronger computing power to keep timely responses. Computing power, as the name implies, refers to the ability of data processing and computing. It exists in all kinds of intelligent hardware devices, ranging from laptops and mobile phones to supercomputers.

At present, computing science is moving from traditional computational and digital simulation to the paradigm composed of high-performance computing, big data, and deep learning. Under such circumstances, computing power also owns different measurements, such as computing speed, algorithm performance, data storage, communication capacity, and computing service capacity. To satisfy the restricted demands, technical explorations are always on the way. For example, the cloud, fog, and edge computing paradigms provide possibilities for the optimization of computing resources, which largely alleviate the computing pressure of the central server. Additionally, leading companies are also devoted to innovative research on hardware and software, which could maximize computing power. Microsoft announces a new supercomputer that enables the training of extremely large AI models. Additionally, it also publishes a new version of the open-source deep learning library for PyTorch, which requires less computing power when training large, distributed models. 

The Metaverse requires stronger computing power to provide timely responses and immersive user experiences. There is an urgent need to continuously develop the techniques related to computing power for future Metaverse.

\ \par \noindent \textbf{d) Artificial Intelligence}
\ \par
Artificial intelligence (AI) is a branch of computer science, which represents the simulation of the information process of human consciousness, thinking, and intelligence \cite{A101}. In recent years, with continuous technical advances, AI techniques have been widely developed and used in both daily life and industrial manufacturing. Generally speaking, the key technologies involved in AI include machine learning \cite{A102,A103}, knowledge graph \cite{A104}, natural language processing \cite{A105,A106}, speech recognition \cite{A107}, computer vision \cite{A108}, etc., which have penetrated into every aspect of our lives. It provides strong support for intelligent applications such as unmanned driving, face recognition, personalized recommendation, and medical image processing \cite{A109}.

In the Metaverse, AI techniques would also contribute a lot to its future development. For example, with the techniques of machine learning, the Metaverse could be easily equipped with strong abilities in data processing and analysis. Techniques of computer vision simulate realistic images and create vivid scenarios and ``avatars'' in the Metaverse. As we all know, the ``avatars'' in the Metaverse are not the simple replication of humans in real life, but also require additional smart functions acting like humans where AI is inseparable. As authors in \cite{A110} conclude, AI would show great significance both in the foundation and development of the Metaverse and would aid a lot in providing immersive user experiences.

\ \par \noindent \textbf{e) Blockchain}
\ \par
Blockchain refers to a decentralized and intelligent distributed ledger platform that helps establish a shareable, trustworthy, and durable mechanism. It is composed of blocks that automatically form a chain according to the time of generation and could only be revised according to strict rules and open protocols \cite{A111}. Instead of relying on other institutions to provide credit evaluation, the Blockchain allows the adoption of technologies such as cryptography and computer science to ensure security and build a corresponding secure shared database. Due to these characteristics, it has huge potential in finance, logistics, and other areas with high-security demands. 

For example, the Blockchain could get rid of the dependence on the third-party and realize direct point-to-point connection in financial fields, which not only reduce costs but also help complete transaction payment quickly \cite{A112}. In the area of smart logistics, the Blockchain could trace the production and delivery process of items and improve the efficiency of supply chain management with lower cost \cite{A113}. In addition, the Blockchain can prove the existence of a file or digital content at a specific time through a hash timestamp, and its characteristics of openness, non-tampering, and traceability provide a complete solution for judicial authentication protection and anti-counterfeiting traceability. Moreover, the Blockchain can simplify the complex multi-level structure in the energy system and reduce the transaction cost of energy as well \cite{A114}.


Blockchain technology has demonstrated its high performance in optimizing big data applications and data circulation as well as data sharing. To the best of our knowledge, the future Metaverse will generate massive amounts of data, and the existing centralized data storage mechanism fails to tackle such overwhelming challenges. Therefore, the storage and processing solutions with Blockchain are expected to show significance in the future Metaverse. In addition, the non-tampering and traceability features of blockchain ensure the authenticity and high quality of data and would provide the basis for the development of Metaverse which also owns specific virtual economic systems \cite{A116}.

\subsubsection{Potential techniques in aspect of space convergence}

\begin{figure}[!ht]
\centering
\includegraphics[width=7.5cm]{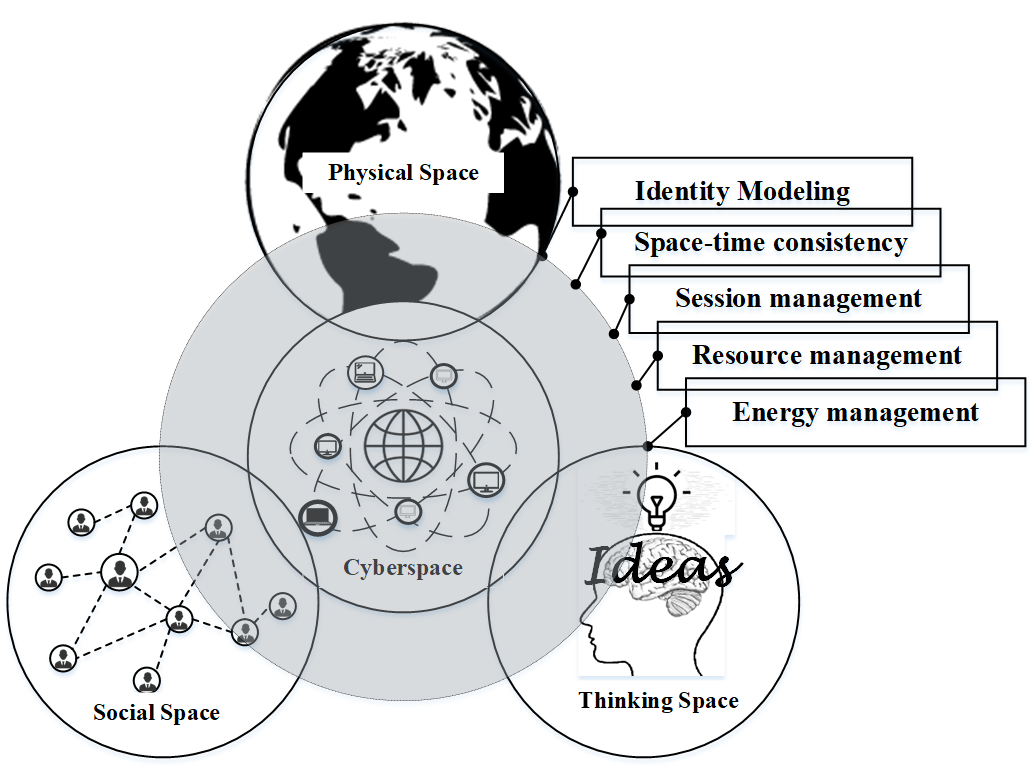}
\caption{Techniques related to space convergence.}
\label{4}
\end{figure}

The convergence among physical, social, thinking, and cyber spaces is promoted by the development of Internet technologies and has gradually formed the CPST hyperspace \cite{A1}. As can be seen in Figure \ref{4}, the techniques in the aspect of space convergence enable it possible to break the spatial boundaries and lay the fundamental foundation for the emergence of the Metaverse. In this section, we will introduce the enabling techniques of space convergence, which would provide possibilities for the future Metaverse.

\ \par \noindent \textbf{a) Identity modeling}
\ \par
As the Metaverse depicts a digitalized world that is similar to the real space, it includes elements such as ``human'', ``animal'', and ``plant'', as well as the economy, education, traveling, rules and laws. Identity modeling plays a significant role in the Metaverse, which will not only help create an identification for given objects but also achieve accurate mapping between the Metaverse and the real world.

Generally, identity modeling can be divided into two categories, ID-based and nID-based identity modeling. As the name implies, ID-based identity modeling depicts the identity by specific codes, like the ID card of humans in the real world. Research on ID-based identity is always on the way. For instance, Masinter et al. propose a Uniform Resource Identifier scheme with a string of characters to identify objects \cite{A48}. Brock et al. put forward the Electronic Product Code (EPC) based on Universal Product Code (UPC), which adopts simple and extensible codes to track objects throughout their life cycle \cite{A49}. All these explorations guide ID-based identity modeling which is also applicable in the future Metaverse.

In addition to ID-based identity modeling with tags assigned by the outside, the objects also have specific attributes themselves, such as natural attributes and social attributes. Ning et al. first proposed the concept of nID which would be used to represent objects when ID does not exist, is untrusted, or is damaged \cite{A50}. He further presented a tree-code model combined with ID and nID information to realize the representation of objects \cite{A51}. Usually, in nID-based identity modeling, the techniques such as semantic ontology are commonly used for representing the attributes, especially the RDF and OWL ontology languages \cite{A52}. For example, Hasemann et al. models embedded devices in the RDF and OWL language and extends this identification method to a variety of hardware scenarios \cite{A53}.

With the help of identity modeling, accurate mapping between the Metaverse and the real world would be achieved and would also support higher applications in the Metaverse such as relationship traceability, safety certification, and unified management.

\ \par \noindent \textbf{b) Space-time consistency}
\ \par
The accurate mapping between the real space and the Metaverse also puts forward high requirements for space-time consistency, to achieve seamless interactions.

There is already some research on spatial-temporal consistency between real space and virtual space. For example, Li et al. proposed a space-time registration model to realize the space and time synchronization between reality and virtuality, which adopted the electronic support measure and unscented Kalman filter to compute the space-time biases and perform time synchronization respectively \cite{A54}. Besides, Zhou et al. proposed a metric to quantify the space-time inconsistency in large-scale distributed virtual environments and verified that this metric can effectively evaluate space-time consistency through Ping-Pong game experiments \cite{A55}. Additionally, Zhong et al. realized the long sequence tracking of objects in a moving object detection system based on the characteristics of space-time consistency \cite{A56}. All these technical explorations could guide for achieving space-time consistency in the future Metaverse. 

The Metaverse depicts a world full of various activities, and space-time consistency is so significant to keep communications and collaboration smooth and successful.

\ \par \noindent \textbf{c) Session management}
\ \par
In the Metaverse, an increasing number of immersive services would be available such as virtual dressing, digital transactions, and space travel. There is also a need to track and monitor humans' activities in real-time to guarantee interactions and communication successfully. In this regard, session management plays an important role. It not only refers to the management of sessions across real life and the Metaverse but also the applications and services that take place in the Metaverse.

Nowadays, many researchers have done related work on session management in virtual space, especially on the Internet, which can be adopted in the future Metaverse. Johnston used a password authentication method based on HTML forms to track user authentication and login sessions in mail order sales websites \cite{A57}. Gutzmann supervised the behavior of users who accessed the network through session management in the HTTP environment, which was realized by the cookie-based ticket mechanism \cite{A58}. Additionally, Poggi et al. used machine learning algorithms and the Markov-chain model to predict the interactive behavior of web users, according to which resources will be reasonably allocated to users in advance \cite{A59}. 

Session management technology mainly includes those technologies that are meaningful to security and privacy protection to a certain extent. With the help of identity authentication, access control, algorithm encryption, and other solutions, it ensures the transparency and security of sessions. These ideas would inspire a lot for the session management in the Metaverse and provide some guidance for further development.

\ \par
\noindent
\textbf{d) Resource  management}
\ \par
When it comes to resources in the Metaverse, we are going to discuss it from two aspects, namely the hardware and software. For example, the supporting hardware resources mainly refer to the core components, including chips, display screens, interactive devices, terminals, etc., while the software may refer to the operating systems, the rendering tools, the computing capabilities, the memory storage, file data, etc. Reasonable resource management could prompt the proper and orderly resource allocation according to dynamic requirements and guarantee a stable and disciplined resource-using environment for the Metaverse.

It is worth noting that there is already some research on resource management. For example, Czajkowski et al. propose a resource management architecture that could solve the problem of resource allocation between different components. This resource management architecture has been verified on a meta-computing system platform, including 15 sites, 330 computers, and 3,600 processors \cite{A63}. Kim et al. build an efficient resource management scheme (ERMS), which allocates the IoT storage resources based on the XML standard, laying a good foundation for the distributed storage and process of data \cite{A64}.

Since the Metaverse is still in its infancy, there is currently no mature mechanism of resource management. However, the existing works related to resource management in IoT or cyberspace could be adopted for reference. In addition, the Metaverse will own some novel resources like ``avatars'' and virtual identities, and there will be additional requirements needed to be satisfied in its resource management, such as the ownership, management, and disposal of multiple virtual identities.

\ \par \noindent \textbf{e) Energy management}
\ \par
As we can imagine, the Metaverse provides immersive applications and services to human beings depending on various types of energy. There may be some novel forms of energy such as thin films for solar cells, in addition to oil, steam, and electricity. It puts forward high requirements for effective energy management.

There already exist some explorations of energy management in real life, which could guide that in the Metaverse. For instance, based on the principle of harvesting energy and saving electricity, Zhang et al. transformed the network energy utility optimization problem into a stochastic problem and proposed a time-scale energy allocation algorithm based on the Lyapuno framework \cite{A65}. Yang et al. modeled the energy optimization problem as the multi-agent reinforcement learning formula and proposed an energy management method based on collaborative multi-agent deep reinforcement learning, which could be used to implement radio block assignment and the transmission power control strategy \cite{A66}. All these techniques would inspire energy management in the Metaverse.

Moreover, there is a need to support the interactions between the Metaverse and the real space, hence energy management may need to overcome the boundaries between virtuality and reality. Considering the specific requirements, the open issues such as the energy transmission between virtual and real space, the consumption calculation, and optimization are needed to be resolved. There is still a long way to explore efficient energy management in the Metaverse.

\subsubsection{Potential techniques in aspect of virtuality and reality interaction}

As we talked about above, the Metaverse is such a paralleled digitalized world while humans are still in the real physical space, hence there is an overwhelming interaction between virtuality and reality. Related technologies are not only a better way to enter the virtual Metaverse, but also the best choice to create immersive scenes in the Metaverse. At present, there have been some relatively mature explorations in virtuality and reality interaction. In this section, we mainly introduce some typical ones which may contribute a lot to the development of the Metaverse.

\ \par \noindent \textbf{a) Virtual Reality (VR)}
\ \par
The concept of VR was first proposed by Jaron Lanier in 1989 when he pointed out that VR reestablished the relationships with the physical world in a new plane \cite{A30}. However, it is worth noting that there would not exert any influence on the subjective world. The VR only helps depict a virtual environment generated by computers, which usually depends on external devices such as helmets and glasses. It could help generate a real-time simulation through multi-sensory channels of taste, smell, vision, sound, touch, etc. so that users are completely immersed in a virtual environment \cite{A29}.

With the help of VR, it is easier to effectively make people believe in the environments they feel and achieve an unprecedented immersive experience. In general, VR techniques in the field of 3D audiovisual modeling, tracking, and interaction show high efficiency. For example, 3D audiovisual technologies can ``puzzle'' the brain, and its high-definition realistic effect makes you believe what you see is what you get \cite{A32}. 3D tracking technologies could track the movements and rotations of the observers to realize precise real-time positioning. It has been adopted in areas such as autonomous driving \cite{A34}. 3D interactive technologies provide possibilities for users to be immersed in the virtual world. Some common tools including intelligent gloves, smart glasses, and helmets are widely adopted for household and commercial purposes \cite{A35}.

To maximize the user's sense of immersion, VR technologies need to break through their limitations such as the strong dependence on devices. In recent years autostereoscopy has become popular since it provides possibilities to display stereoscopic images based on the theories of the parallax barrier, lenticular lens, and light field, rather than depending on any specific devices \cite{A33}. This is a great step forward to the immersive user experience and the future Metaverse, as there is no strict dependence on complicated helmets and glasses, which would largely improve the feelings of users to some extent.

\ \par \noindent \textbf{b) Augmented Reality (AR)}
\ \par
AR is a technology that skillfully blends virtual information with the real world, and it adopts a variety of techniques, including multimedia, 3D modeling, real-time tracking and registration, intelligent interaction, sensing technologies, etc. \cite{A31}. Different from VR technologies, AR emphasizes the extension of the real world and helps enhance the objects residing in the real physical space by adding multiple sensory modalities \cite{A36}.

AR needs the support of a complete set of hardware devices, such as processors, displays, sensors, and input devices. In these components, we could adopt modern mobile computing devices such as smartphones and tablets to act as processors to provide powerful processing and computing capabilities. As for displays, there are some common tools, among which the head-mounted display (HMD) is one kind of display device worn on the forehead, like a helmet. In such Modern HMDs, sensors are typically used for precise monitoring, enabling the system to align virtual information with the physical world and adjust accordingly based on the user's head movements \cite{A37}. Sensors are so significant in AR techniques because the capture of the surrounding real environment still needs to be realized via them, for instance, accelerators, GPS, infrared sensors, and so on. The input devices involved in AR technologies are mainly cameras or webcams. Some even require speech recognition systems that translate what the user says into instructions the computer can understand, and gesture recognition systems that interpret body movements through visual detection or sensors. Giant IoT companies are devoted to the research and development of some advanced AR equipment, for example, the AR glass ``Project Iris'' from Google, and ``Shield AR'' from Vuzix.

VR and AR are a pair of similar concepts, but there are still some differences between them. In VR, the users' perception of reality completely depends on virtual information, while in AR, users gain additional computer-generated information in addition to the data collected in real life, thereby enhancing their perception of reality \cite{A38}. It is worth noting that they are all beneficial technologies to improve users' immersion, and will be key technologies of the future Metaverse.

\ \par \noindent \textbf{c) Mixed Reality (MR)}
\ \par
MR is also another technique that helps us get rid of the restrictions of the screen and improves the users' immersion to a large extent. Rather than creating a completely virtual scene, MR focuses more on the instinctual interactions between the real world and the digitalized space, which is like AR techniques we discussed above \cite{A39}. It could be regarded as a hybrid of AR and VR, where a transition between VR and AR could be achieved simultaneously.

\begin{table}[!ht]
\centering
\caption{The differences between VR, AR, and MR.}
\label{tab2}
\begin{tabular}{ccc}
\toprule
\toprule
Type  &   Description  &   Characteristic  \\
\midrule
VR  & \multicolumn{1}{m{3.5cm}}{In the VR world, everything is virtual and totally created by various techniques.} &   \multicolumn{1}{m{2cm}}{Virtuality} \\
\midrule
AR &   \multicolumn{1}{m{3.5cm}}{The AR technique creates visual things, and then overlies them into the real world.}   & \multicolumn{1}{m{2cm}}{Virtuality and Reality, from virtuality to reality} \\
\midrule
MR   & \multicolumn{1}{m{3.5cm}}{The MR technique visualizes the real things, and then overlies them into the virtual world. It needs to keep instant message with the real world. } & \multicolumn{1}{m{2cm}}{Virtuality and Reality, from reality to virtuality} \\
\bottomrule
\bottomrule
\end{tabular}
\end{table}
MR needs to acquire real-time access to effective information about real objects, achieve its digital modeling, and then realize the co-existence and interaction between the real and visual objects. To better distinguish between VR, AR, and MR, we make a comparison as shown in Table \ref{tab2}.

Generally, MR techniques are more complex because real objects need to be virtualized first. During the process of virtualization, it needs a camera to scan objects for 3D reconstruction. We all know that the picture captured by the camera is two-dimensional, that is, the picture is flat, and the depth information is lost, so it is necessary to reconstruct and generate a virtual 3D object, which we call a real virtualized object. This kind of MR technique is popular in our daily life. For example, the AR filters in social media are providing MR user experiences. In addition, MR techniques also show huge possibilities in areas of education, entertainment, remote working, etc. It will become mainstream in the future Metaverse.

\begin{figure*}[!ht]
\centering
\includegraphics[width=16cm]{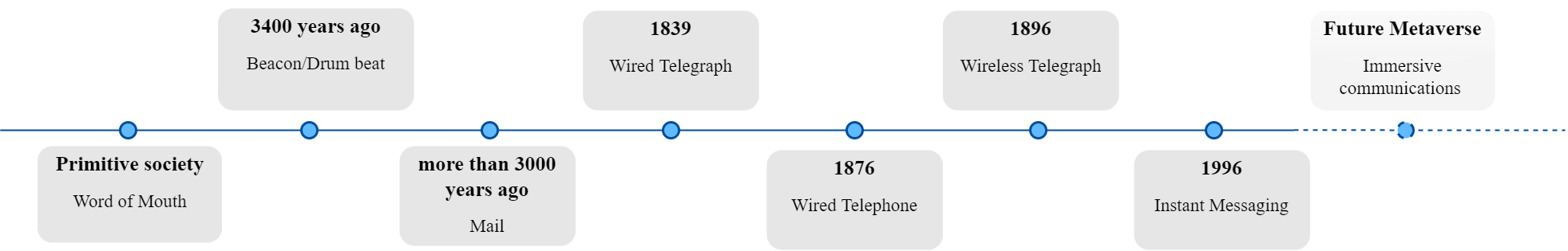}
\caption{The development of human-centered communication.}
\label{fig_4}
\end{figure*}

\ \par \noindent \textbf{d) Brain-Computer Interface (BCI)}
\ \par
BCI sometimes called ``brain port'', is a direct connection path established between the brain and external equipment \cite{A40}. In general, common BCIs can be divided into unidirectional BCIs and bidirectional BCIs, according to the direction of instruction transmission. As the name suggests, in the case of unidirectional BCIs, the computer either accepts commands from the brain or sends signals to the brain but fails to send and receive signals at the same time, while the bidirectional BCIs allow bidirectional information to be exchanged between the brain and external devices simultaneously.

Through the connection between the brain and the computer, people can freely acquire information, socialize with each other, and even obtain such sensory experiences in the virtual world as taste, touch, etc. \cite{A41}. Compared with ``traditional'' media which only provide two-dimensional sensory experiences such as audio and video, BCIs can bring a revolutionary experience to the Metaverse.

Since 2019, Neuralink, Musk's BCI company, has repeatedly set off a wave of public opinion with its advances in BCIs \cite{A43}. It first announced its BCI system on July 17, 2019. On August 29, 2020, Elon Musk showed Neuralink brain implant working in a pig and successfully reading its brain activities \cite{A23}. This is a big step forward as it implies the possibilities in the future for further connections between human brains and computers, which may be one of the hot topics in the future Metaverse.

The emergence of BCIs enhances the interaction between reality and virtual space. In the future Metaverse, it may be possible to control objects in the Metaverse with brain waves to a certain extent, and users can freely move every part of the body according to their wishes. It will no longer need rigidly preset movements, and users can interact with the virtual world as much as possible \cite{A42}.

\ \par \noindent \textbf{e) Game Engine}
\ \par
When it comes to the interaction between virtuality and reality, the gaming industry is one of the first industries to be noticed. Especially in recent years, the popularity of somatosensory games and interactive projection games has brought the gaming industry into a new stage. Under such circumstances, the game engine also emerges gradually, which provides software platforms for game designers to develop various video games. 

The game engine is composed of some editable systems or applications that could allow game designers to develop games much more efficiently and conveniently without starting from scratch \cite{A44}. To some extent, the game engine could be regarded as a predefined set of codes for certain games that the machine could be understood. As authors in \cite{A45} mentioned, the game engine usually includes parts of the rending engine, physics engine, collision detection system, sound effects, script engine, computer animation, network engine, scene management, etc. With the combination of the above technologies, it could integrate different game resources, such as images, sounds, and animations, and design according to the requirements.

Nowadays, the game engine is not only born for games but also can support the creation of interactive high-fidelity graphics and environments. Since the gaming industry is one of the first hot fields in the Metaverse, the game engine also serves as a significant method to realize the Metaverse. In 2020, the company named MetaVRse announced its MetaVRse engine, the new 3D game engine for non-coders \cite{A46}. It enables the users to create immersive experiences in the no-code movement, which is a great step toward the future Metaverse.

\ \par \noindent \textbf{f) 3D modeling}
\ \par
Generally speaking, 3D modeling is constructing a mathematical model of any given object by replicating the parameters of its surface, such as the edges, vertices, polygons, etc. The products created by 3D modeling techniques such as 3D rendering and 3D computer graphics are also named as 3D models. 

3D modeling plays a significant role in education, entertainment, healthcare, etc. For example, 3D modeling can build a virtual museum where people could enjoy all kinds of exquisite artwork without leaving home \cite{A60}. In the area of healthcare, Winzenrieth established a 3D model based on dual-energy x-ray absorptiometry, which allowed a more precise assessment of the drug's therapeutic effect \cite{A61}. Additionally, 3D modeling could also improve the success rate of surgery, and a novel 3D modeling tool for virtual nose surgery has huge potential to help surgeons model potential surgical maneuvers and minimize the complications \cite{A62}. The popular holographic projection is one kind of 3D modeling technique, which originally refers to the technology of recording and reproducing the real 3D image of an object based on the principle of interference. For example, with the help of holographic projection, the late Teresa Teng could ``stand'' on the stage singing and providing a visual feast.

3D modeling technologies hold significant roles in the construction and development of the Metaverse. They help replicate the architectural style of buildings in real life and make the scenarios more realistic. The 3D modeling technologies could be regarded as the foundations of the Metaverse and serve as the important bridge keeping the accurate mapping between the Metaverse and real life.

\ \par \noindent \textbf{g) Real-time rendering}
\ \par
Real-time rendering is a branch of computer graphics that studies how to create and analyze real-time images \cite{A73}. It could quickly render a constantly changing 3D environment and create the illusion of movement. Real-time rendering takes advantage of the ``persistence of vision'' feature of human eyes, which have a short pause in the observed scenes. Hence, we humans could not notice the phenomenon of frame switching as long as the switching speed is fast enough. It looks like a real-time changing animation. At present, the techniques of real-time rendering are mainly used in the graphic design of video games and the production of movies.

Compared with the production of traditional animation, real-time rendering focuses on interactivity and real-time \cite{A74}. Generally, the scene is optimized to improve the calculation speed and reduce the time delay. For users, any operation such as finger swiping across the screen, mouse click, keyboard input, etc., will cause the screen to recalculate. Meanwhile, the users need to get real-time feedback after the operation. Therefore, real-time calculation and response are significant for a better user experience.

However, the quality of real-time rendering is deeply limited by hardware, which drives the industry to focus more on technical innovations. With the improvement of GPU performance, the speed of real-time computing is faster, and the accuracy of image computing has become much higher. For example, NVIDIA launched RTX Real-Time Ray Tracing technology in 2018, which was possible to provide more realistic images \cite{A75}.

Since the Metaverse is a digitalized world, most of the scenes are virtualized and need to be rendered in real-time. With the development of real-time rendering technologies, the Metaverse is possible to provide immersive user experiences, coupled with strong interactivity and in-time response.

\subsubsection{Potential techniques in aspect of human-centered communication}

The development of human-centered communication not only drives the emergence and prosperity of the Metaverse but also puts forward higher requirements for Metaverse development. As can be seen in Figure \ref{fig_4}, human-centered communication has gone through several stages, from primitive communication relying on body movements or languages to the emergence of wired and wireless communication and instant messaging in the 20th century. Great changes have taken place in the aspect of human-centered communication, especially in recent years, the development of various social media like Instagram, MSN, Facebook, WeChat, etc., have penetrated every corner of modern life, which greatly breaks through physical boundaries and provides much convenience for communication.

However, at this stage, communication still relies on physical media, such as mobile phones, computers, and other electronic devices. Users are still faced with ``screens'' and could not achieve a more immersive user experience. While the Metaverse serves as a higher-level living space, it gives birth to the new communication of advanced immersion and feelings. By enabling multi-dimensional and multi-sensory communication methods, users are allowed to communicate and establish immersive social contacts with each other, further breaking the spatial-temporal limitations. Usually, the communication methods in the Metaverse include traditional audio-visual communication through text, voice, image, etc., and some higher-dimensional communication ways such as simulating touch, taste, and other senses.

Under such circumstances, it is necessary to perceive and process complex and multi-dimensional information in the Metaverse to keep successful communication. Hence techniques related to 3D communication methods, social networks analysis, social aware computing, cognitive computing, swarm intelligence, and affective computing are becoming vital, which would help a lot during human-centered communication. In this section, we will give a further introduction to the detailed techniques.

\ \par \noindent \textbf{a) 3D communication methods}
\ \par
The Metaverse should provide users with more free and open communication scenarios, to improve users' communication experience and efficiency. In addition to traditional audio-visual communication ways, more immersive ways of communication like signals, tastes, and feelings are expected. As authors in \cite{A67} argue, the Metaverse allows users to communicate seamlessly with digital artifacts via different approaches, for instance, the XR systems with stereoscopic displays, spatial or binaural audio with soundscape constructions, handheld input devices, etc.

Specifically, handheld input devices play specific roles during human-centered communication in the Metaverse. For example, the motion controllers which allow people to touch, grab, and manipulate visual objects could support active interaction and communication \cite{A68}. Additionally, multi-sensory communication overcomes the restrictions of ``screens'', where humans could communicate their feelings with each other face to face, even if they are still in different places in real space. It gets rid of the shackles of screens or other intermediate media, and largely improves users' immersive experience during communication.

\ \par \noindent \textbf{b) Social networks analysis}
\ \par
The social network is a typical complex network, which is defined as a social structure composed of various nodes and edges. Nodes usually stand for individuals or organizations, while edges usually refer to social relationships. Social networks contain massive amounts of information, reflecting the group interaction behaviors and social relationships among users \cite{A69}. It shows unique characteristics due to its sociality represented in the graph structure, such as centrality, betweenness, modularity, etc. Via social network analysis, more hidden information could be obtained \cite{A70}. There are two kinds of social network structures, the small world networks, and the scale-free networks. 

In small-world networks, two important properties named network average path length, and average aggregation coefficient count a lot \cite{A71}. The Network average path length refers to the average distance between nodes in the network, while the average aggregation coefficient refers to the probability of any node whose neighbors are also neighbors. By analyzing these parameters, it is possible to predicate the hidden relationships between different nodes.

The scale-free features come from the study of complex networks, that is, the degree to in the network obeys the power-law distribution \cite{A72}. In social networks, it means that most nodes have a small number of edges connected, while a few nodes have a large number of edges connected. That is, a few people have complex social relationships, while most people have simple social relationships. According to the research, well-known social networks such as Flickr, YouTube, and LiveJournal all have small-world phenomena and scale-free characteristics \cite{A97}.

Analyzing the graph structure and social characteristics in social networks, allows us to study the laws existing in the social networks, recur and deduct the social features. At present, social network analysis is mostly used for social recommendation \cite{A98}, social influence modeling \cite{A99}, user behavior prediction, etc. \cite{A100}. Since the relationships and interactions in the Metaverse are not limited to humans, ``avatar'', but humans and ``avatar'', the social network analysis techniques would contribute a lot to the Metaverse, especially in the predication, deduction, and management of the complicated social relationships and behaviors, etc.

\ \par \noindent \textbf{c) Social aware computing}
\ \par
To our best knowledge, the continuous developing of pervasive computing largely improves the ability to data acquisition and process. Combined with the progress in social computing, a new paradigm of social aware computing emerges. Social aware computing refers to the process of real-time perception and social behavior recognition, which could analyze and mine social characteristics and specialization, further supporting interaction, communication, and cooperation in the community \cite{A76}. Since the Metaverse could be regarded as a world parallel to the real world, which has its special structure and social system, it is equally important to carry on social aware computing in the Metaverse.

The techniques of social aware computing mainly include the following aspects like large-scale pervasive sensing, activity and interaction analysis, social interaction support, software framework and methodology, and applications. Based upon these techniques, it is possible to acquire real-time data via sensing devices, execute analysis of social behavior, and provide auxiliary suggestions for social activities. At present, social aware computing has been widely used in areas of network communication and smart traffic \cite{A79}. For example, Liu designs a congestion control scheme based on social aware computing in Delay Tolerant Networks \cite{A77}. Zhang adopts social characteristic computing in urban informatics, and his proposed optimal scheme largely improves the efficiency of content dispatch \cite{A78}.

As the Metaverse aims to depict an immersive digital world with complicated social systems and mechanisms, it puts forward higher requirements in context awareness and social computing. The techniques of social aware computing make it possible to understand complicated environments well and provide much more intelligent services and user experiences.

\ \par \noindent \textbf{d) Cognitive computing}
\ \par
Cognitive computing refers to the abilities of intelligent systems that could simulate the cognition science of humans, especially the way that the human brain works. With the help of techniques such as information analysis, natural language processing, machine learning, etc., cognitive computing is possible to synthesize data from surroundings and make the most intelligent decisions as required. It attempts to solve the inaccuracy and uncertainty in biological systems and realizes the process of perception, memory, learning, language, thinking, and problem-solving to different degrees \cite{A80}. By training and learning from large amounts of data, cognitive computing systems could improve the way of pattern recognition and data processing, to provide more possibilities for problem prediction and solution modeling. 

Cognitive computing aims to infinitely approach human intelligence to better deal with complex problems of learning, reasoning, deduction, and so forth. For example, IBM Watson is such an expert system that it could help doctors treat their patients by providing suggestions based on its knowledge system during the medical process \cite{A81}. However, it is worth noting that specific abilities are needed to achieve cognitive computing, such as adaptability, interactivity, state iteration, and context understanding. Adaptability ensures the ability to adjust to dynamic data and environment changes, and interactivity represents the capacity of interacting with external humans or devices. State iteration enables the identification of unsure problems by asking questions or extracting additional information, and context understanding emphasizes the capability of better understanding, identifying, and mining contextual data. All these key attributes make cognitive computing successful.

In the Metaverse, there would exist simulated systems that act like humans, for example, the ``avatar'' with their own ``ideas'', ``thoughts'', ``intelligence'', etc. Cognitive computing is quite common and expected. The man-like intelligent systems could largely augment human intelligence and are possible to deal with more complicated issues. Cognitive computing will strengthen Metaverse with efficient data processing speed, abundant expert knowledge, and provide the most appropriate decisions as expected.

\ \par \noindent \textbf{e) Swarm Intelligence}
\ \par
Swarm Intelligence is a type of method that is inspired by the intelligent behaviors of ant colonies or bees in nature, who would like to cooperate and communicate together according to specific rules to realize the final aim \cite{A83}. The concept was first proposed in 1989 by Gerardo Beni and Jing Wang and was adopted in the cellular robotic systems, in which a collection of robots worked together to achieve the goal tasks, even each only with limited processing abilities \cite{A82}. 

Generally, in cyberspace and IoT, swarm intelligence has shown potential in managing many hardware devices and achieving complex functions. For example, Luo proposes a fault-tolerance algorithm based on practical swarm optimization for IoT, to overcome the fault-tolerance routing issue \cite{A85}. Zedadra overviews the swarm intelligence-based algorithms and points out the possibilities to be applied in IoT systems, including node localization, optimization control, medical care, etc \cite{A87}. In \cite{A88}, the authors also design an approach based on swarm intelligence for traffic light scheduling, which has brought huge profits in practical applications. 

In the framework of swarm intelligence, it usually includes three parts: perception participants, network layer and end users \cite{A86}. The perceptual participants refer to intelligent robots and group users, and the network layer is responsible for data storage, transmission, and process. The end users are those who send requirements and look forward to feedback from different perceptual participants and network layers. They cooperate and work for the best results. 

Considering its characteristics, swarm intelligence is more significant for supporting distributed social structures. As Chen concluded, the Metaverse is a digital world where entities could cooperate and tackle significant difficulties together. It is less limited by space and time with more decentralized social characteristics \cite{A84}. Therefore, swarm intelligence could play an important role in the Metaverse, and efficiently guide further system construction and social management.

\section{Open issues and challenges of technologies in the Metaverse}

Since the Metaverse is still in its infancy, there is a long way to establish a comprehensive and mature technical framework at once. Some open issues and challenges of technologies need to be considered for further development in the future Metaverse.

\subsection{The balance between virtuality and reality}

The Metaverse is a digitalized virtual world with strong interaction between virtuality and reality. Some disputes about the interaction between virtuality and reality emerge. Whether virtuality would completely replace reality or reversely drive the development of reality. How do humans switch freely between virtuality and reality and achieve a balance. How to deal with the conflicts and coordination between virtuality and reality. These are all challenges needed to be considered in the future development of the Metaverse.

\subsection{The fusion between different techniques}

The Metaverse relies on various techniques to support further applications. Hence, it may need seamless fusion, coordination, and collaboration between different technologies to address various challenges. It is necessary to better manage the superposition, compatibility, and integration between multiple technologies, to keep the technical system stable and reliable.

\subsection{The gap between theoretical research and practical applications}

Although the Metaverse has become so popular in recent days, most of its research is still in its infancy and stays in the primary stage of theoretical explorations. The huge gap between theoretical research and practical applications is one of the challenges in the Metaverse. It is essential to consider the development of infrastructures, the imbalance of technical resources, the lack of the recognized industry standard, the difficulties in large-scale production, etc. In addition, it also needs to explore the most appropriate forms of business models and interactive ways used in the Metaverse.

\subsection{The resource and energy management}

The problems related to resource and energy management are also open issues. For example, how to maximize the utilization of resources in the Metaverse; how to deal with the problem of energy consumption in the Metaverse; how to solve the contradiction between technological development and energy consumption and resource imbalance. They would become dominant challenges in the technology development of the Metaverse.

\subsection{The security and privacy protection}

Since the Metaverse is a parallel and digitalized world, most personal information of users is at risk of leakage and invasion. It is urgent to establish a safe and comprehensive security and privacy protection mechanism in the Metaverse in case of any risks, for instance, adopting resolutions to strengthen identity authentication and security management, etc.

\section{Conclusions}

The Metaverse has stirred up many hot topics in both academia and industry, and leads to novel explorations. Different from other works, we introduce the Metaverse from a new technology perspective, including its essence, corresponding technical framework, and possible technical challenges in the future. We especially conclude four pillars of the Metaverse, named ubiquitous connections, space convergence, virtuality and reality interaction, and human-centered communication, and establish the corresponding technical framework. Moreover, we analyze some open issues and challenges of the Metaverse in its technical aspect. The Metaverse depicts a digitalized world with immersive user experience, but most of its research is at the theoretical level, and there is still a long way to go in the future.

\bibliographystyle{elsarticle-num}

\bibliography{myreference}

\begin{thebibliography}{100}
\expandafter\ifx\csname url\endcsname\relax
  \def\url#1{\texttt{#1}}\fi
\expandafter\ifx\csname urlprefix\endcsname\relax\def\urlprefix{URL }\fi
\expandafter\ifx\csname href\endcsname\relax
  \def\href#1#2{#2} \def\path#1{#1}\fi

\bibitem{A1}
H.~Ning, X.~Ye, M.~A. Bouras, D.~Wei, M.~Daneshmand, General cyberspace:
  Cyberspace and cyber-enabled spaces, IEEE Internet of Things Journal 5~(3)
  (2018) 1843--1856.

\bibitem{A2}
H.~Ning, H.~Wang, Y.~Lin, W.~Wang, S.~Dhelim, F.~Farha, J.~Ding, M.~Daneshmand,
  A survey on metaverse: the state-of-the-art, technologies, applications, and
  challenges, arXiv preprint arXiv:2111.09673.

\bibitem{A3}
C.~Jaynes, W.~B. Seales, K.~Calvert, Z.~Fei, J.~Griffioen, The metaverse: a
  networked collection of inexpensive, self-configuring, immersive
  environments, in: Proceedings of the workshop on Virtual environments 2003,
  2003, pp. 115--124.

\bibitem{A4}
S.~Spielberg, A.~Silvestri, Z.~Penn, E.~Cline, D.~De~Line, Ready player one,
  Warner Bros USA, 2018.

\bibitem{A5}
H.~Duan, J.~Li, S.~Fan, Z.~Lin, X.~Wu, W.~Cai, Metaverse for social good: A
  university campus prototype, in: Proceedings of the 29th ACM International
  Conference on Multimedia, 2021, pp. 153--161.

\bibitem{A6}
H.~Noon, Something about the six core technologies in the metaverse, Available
  online:
  \url{https://ramaonhealthcare.com/something-about-the-six-core-technologies-in-the-metaverse/}
  (accessed on 30 September 2022).

\bibitem{A7}
L.~Lee, T.~Braud, P.~Zhou, L.~Wang, D.~Xu, Z.~Lin, A.~Kumar, C.~Bermejo,
  P.~Hui, All one needs to know about metaverse: A complete survey on
  technological singularity, virtual ecosystem, and research agenda, arXiv
  preprint arXiv:2110.05352.

\bibitem{A8}
J.~D.~N. Dionisio, W.~G.~B. III, R.~Gilbert, 3d virtual worlds and the
  metaverse: Current status and future possibilities, ACM Computing Surveys
  (CSUR) 45~(3) (2013) 1--38.

\bibitem{A117}
F.~A. Wolf, Parallel universes, Simon and Schuster, 1988.

\bibitem{A119}
S.~McLeod, Maslow's hierarchy of needs, Simply psychology 1~(1-18).

\bibitem{A14}
W.~Chris, Ubiquitous connectivity is here, and it changes everything, Available
  online:
  \url{https://www.forbes.com/sites/forbestechcouncil/2021/08/25/ubiquitous-connectivity-is-here-and-it-changes-everything/?sh=1361147e7cb1}
  (accessed on 30 November 2021).

\bibitem{A9}
H.~Ning, Unit and ubiquitous internet of things, CRC press, 2013.

\bibitem{A10}
F.~M. Schaf, S.~Paladini, C.~E. Pereira, 3d autosyslab prototype, in:
  Proceedings of the 2012 IEEE Global Engineering Education Conference
  (EDUCON), IEEE, 2012, pp. 1--9.

\bibitem{A11}
Y.~Han, D.~Niyato, C.~Leung, C.~Miao, D.~I. Kim, A dynamic resource allocation
  framework for synchronizing metaverse with iot service and data, arXiv
  preprint arXiv:2111.00431.

\bibitem{A12}
N.~Mohammadi, J.~E. Taylor, Smart city digital twins, in: 2017 IEEE Symposium
  Series on Computational Intelligence (SSCI), 2017.

\bibitem{A13}
Q.~Lu, A.~K. Parlikad, P.~Woodall, G.~Don~Ranasinghe, X.~Xie, Z.~Liang,
  E.~Konstantinou, J.~Heaton, J.~Schooling, Developing a digital twin at
  building and city levels: Case study of west cambridge campus, Journal of
  Management in Engineering 36~(3) (2020) 05020004.

\bibitem{A15}
R.~Baheti, H.~Gill, Cyber-physical systems, The impact of control technology
  12~(1) (2011) 161--166.

\bibitem{A16}
Y.~Zhou, F.~R. Yu, J.~Chen, Y.~Kuo, Cyber-physical-social systems: A
  state-of-the-art survey, challenges and opportunities, IEEE Communications
  Surveys \& Tutorials 22~(1) (2019) 389--425.

\bibitem{A17}
H.~Ning, H.~Liu, Cyber-physical-social-thinking space based science and
  technology framework for the internet of things, Science China Information
  Sciences 58~(3) (2015) 1--19.

\bibitem{A18}
A.~Khanna, R.~Anand, Iot based smart parking system, in: 2016 International
  Conference on Internet of Things and Applications (IOTA), IEEE, 2016, pp.
  266--270.

\bibitem{A19}
F.~Tao, Q.~Qi, L.~Wang, A.~Nee, Digital twins and cyber--physical systems
  toward smart manufacturing and industry 4.0: Correlation and comparison,
  Engineering 5~(4) (2019) 653--661.

\bibitem{A20}
P.~A. Grabowicz, J.~J. Ramasco, E.~Moro, J.~M. Pujol, V.~M. Eguiluz, Social
  features of online networks: The strength of intermediary ties in online
  social media, Plos One 7~(1) (2012) e29358.

\bibitem{A22}
S.~Steinert, O.~Friedrich, Wired emotions: Ethical issues of affective
  brain--computer interfaces, Science and engineering ethics 26~(1) (2020)
  351--367.

\bibitem{A21}
E.~Musk, et~al., An integrated brain-machine interface platform with thousands
  of channels, Journal of medical Internet research 21~(10) (2019) e16194.

\bibitem{A23}
T.~Lewis, Elon musk's pig-brain implant is still a long way from `solving
  paralysis', Available online:
  \url{https://www.scientificamerican.com/article/elon-musks-pig-brain-implant-is-still-a-long-way-from-solving-paralysis/}
  (accessed on 30 November 2021).

\bibitem{A25}
X.~Hu, R.~Su, L.~He, The design and implementation of the 3d educational game
  based on vr headsets, in: 2016 international Symposium on educational
  technology (ISET), IEEE, 2016, pp. 53--56.

\bibitem{A24}
R.~Epp, D.~Lin, C.~Bezemer, An empirical study of trends of popular virtual
  reality games and their complaints, IEEE Transactions on Games 13~(3) (2021)
  275--286.

\bibitem{A26}
Y.~Liu, S.~Wu, Q.~Xu, H.~Liu, Holographic projection technology in the field of
  digital media art, Wireless Communications and Mobile Computing 2021.

\bibitem{A27}
Z.~Guo, Z.~Wang, X.~Jin, ``avatar to person'' (atp) virtual human social
  ability enhanced system for disabled people, Wireless Communications and
  Mobile Computing 2021.

\bibitem{A28}
W.~Chang, H.~Shin, Virtual experience in the performing arts: K-live hologram
  music concerts, Popular Entertainment Studies 10~(1-2) (2020) 34--50.

\bibitem{A89}
G.~Sagl, B.~Resch, Mobile Phones as Ubiquitous Social and Environmental
  Geo-Sensors, Mobile Phones as Ubiquitous Social and Environmental
  Geo-Sensors, 2015.

\bibitem{A90}
Q.~Li, W.~Huangfu, F.~Farha, T.~Zhu, S.~Yang, L.~Chen, H.~Ning,
  \href{https://www.sciencedirect.com/science/article/pii/S0167739X19325361}{Multi-resident
  type recognition based on ambient sensors activity}, Future Generation
  Computer Systems 112 (2020) 108--115.
\newblock \href
  {http://dx.doi.org/https://doi.org/10.1016/j.future.2020.04.039}
  {\path{doi:https://doi.org/10.1016/j.future.2020.04.039}}.
\newline\urlprefix\url{https://www.sciencedirect.com/science/article/pii/S0167739X19325361}

\bibitem{A91}
K.~Chang, Bluetooth: a viable solution for iot?[industry perspectives], IEEE
  Wireless Communications 21~(6) (2014) 6--7.

\bibitem{A92}
J.~Yang, Z.~Han, H.~Jiang, L.~Xie, Device-free occupant activity sensing using
  wifi-enabled iot devices for smart homes, IEEE Internet of Things Journal PP
  (2018) 1--1.

\bibitem{A93}
J.~Huang, F.~Ruan, S.~Ming, X.~Yang, J.~Zhang, Analysis of orthogonal frequency
  division multiplexing (ofdm) technology in wireless communication process,
  in: 2016 10th IEEE International Conference on Anti-counterfeiting, Security,
  and Identification (ASID), 2016.

\bibitem{A94}
L.~Chettri, R.~Bera, A comprehensive survey on internet of things (iot) toward
  5g wireless systems, IEEE Internet of Things Journal 7~(1) (2020) 16--32.

\bibitem{A95}
D.~C. Nguyen, M.~Ding, P.~N. Pathirana, A.~Seneviratne, J.~Li, D.~Niyato,
  O.~Dobre, H.~V. Poor, 6g internet of things: A comprehensive survey, IEEE
  Internet of Things Journal.

\bibitem{A101}
P.~H. Winston, Artificial intelligence, Addison-Wesley Longman Publishing Co.,
  Inc., 1992.

\bibitem{A102}
M.~I. Jordan, T.~M. Mitchell, Machine learning: Trends, perspectives, and
  prospects, Science 349~(6245) (2015) 255--260.

\bibitem{A103}
B.~Mahesh, Machine learning algorithms-a review, International Journal of
  Science and Research (IJSR).[Internet] 9 (2020) 381--386.

\bibitem{A104}
J.~Pujara, H.~Miao, L.~Getoor, W.~Cohen, Knowledge graph identification, in:
  International semantic web conference, Springer, 2013, pp. 542--557.

\bibitem{A105}
J.~Hirschberg, C.~D. Manning, Advances in natural language processing, Science
  349~(6245) (2015) 261--266.

\bibitem{A106}
C.~Wu, X.~Li, Y.~Guo, J.~Wang, Z.~Ren, M.~Wang, Z.~Yang, Natural language
  processing for smart construction: Current status and future directions,
  Automation in Construction 134 (2022) 104059.

\bibitem{A107}
A.~V. Haridas, R.~Marimuthu, V.~G. Sivakumar, A critical review and analysis on
  techniques of speech recognition: The road ahead, International Journal of
  Knowledge-Based and Intelligent Engineering Systems 22~(1) (2018) 39--57.

\bibitem{A108}
X.~Feng, Y.~Jiang, X.~Yang, M.~Du, X.~Li, Computer vision algorithms and
  hardware implementations: A survey, Integration 69 (2019) 309--320.

\bibitem{A109}
E.~Soegoto, R.~Utami, Y.~Hermawan, Influence of artificial intelligence in
  automotive industry, in: Journal of Physics: Conference Series, Vol. 1402,
  IOP Publishing, 2019, p. 066081.

\bibitem{A110}
T.~Huynhthe, Q.~Pham, X.~Pham, et~al., Artificial intelligence for the
  metaverse: A survey, arXiv preprint arXiv:2202.10336.

\bibitem{A111}
W.~Viriyasitavat, D.~Hoonsopon, Blockchain characteristics and consensus in
  modern business processes, Journal of Industrial Information Integration 13
  (2019) 32--39.

\bibitem{A112}
V.~Chang, P.~Baudier, H.~Zhang, Q.~Xu, J.~Zhang, M.~Arami, How blockchain can
  impact financial services--the overview, challenges and recommendations from
  expert interviewees, Technological forecasting and social change 158 (2020)
  120166.

\bibitem{A113}
E.~Tijan, S.~Aksentijevi{\'c}, K.~Ivani{\'c}, M.~Jardas, Blockchain technology
  implementation in logistics, Sustainability 11~(4) (2019) 1185.

\bibitem{A114}
M.~Andoni, V.~Robu, D.~Flynn, S.~Abram, D.~Geach, D.~Jenkins, P.~McCallum,
  A.~Peacock, Blockchain technology in the energy sector: A systematic review
  of challenges and opportunities, Renewable and Sustainable Energy Reviews 100
  (2019) 143--174.

\bibitem{A116}
H.~Jeon, H.~Youn, S.~Ko, T.~Kim, Blockchain and ai meet in the metaverse,
  Advances in the Convergence of Blockchain and Artificial Intelligence (2022)
  73.

\bibitem{A48}
L.~Masinter, T.~Bernerslee, R.~T. Fielding, Uniform resource identifier (uri):
  Generic syntax, Network Working Group: Fremont, CA, USA.

\bibitem{A49}
D.~L. Brock, The electronic product code (epc), Auto-ID Center White Paper
  MIT-AUTOID-WH-002 (2001) 1--21.

\bibitem{A50}
H.~Ning, S.~Hu, W.~He, Q.~Xu, H.~Liu, W.~Chen, nid-based internet of things and
  its application in airport aviation risk management, Chinese Journal of
  Electronics 21~(2) (2012) 209--214.

\bibitem{A51}
H.~Ning, Y.~Fu, S.~Hu, H.~Liu, Tree-code modeling and addressing for non-id
  physical objects in the internet of things, Telecommunication Systems 58~(3)
  (2015) 195--204.

\bibitem{A52}
I.~Szilagyi, P.~Wira, Ontologies and semantic web for the internet of things-a
  survey, in: IECON 2016-42nd Annual Conference of the IEEE Industrial
  Electronics Society, IEEE, 2016, pp. 6949--6954.

\bibitem{A53}
H.~Hasemann, A.~Kr{\"o}ller, M.~Pagel, The wiselib tuplestore: a modular rdf
  database for the internet of things, arXiv preprint arXiv:1402.7228.

\bibitem{A54}
W.~Li, H.~Leung, Y.~Zhou, Space-time registration of radar and esm using
  unscented kalman filter, IEEE Transactions on Aerospace and Electronic
  Systems 40~(3) (2004) 824--836.

\bibitem{A55}
S.~Zhou, W.~Cai, B.-S. Lee, S.~J. Turner, Time-space consistency in large-scale
  distributed virtual environments, ACM Transactions on Modeling and Computer
  Simulation (TOMACS) 14~(1) (2004) 31--47.

\bibitem{A56}
D.~Zhong, S.~Chang, Long-term moving object segmentation and tracking using
  spatio-temporal consistency, in: Proceedings 2001 International Conference on
  Image Processing (Cat. No. 01CH37205), Vol.~2, IEEE, 2001, pp. 57--60.

\bibitem{A57}
P.~Johnston, Authentication and session management on the web, Retrieved
  December 13 (2004) 2009.

\bibitem{A58}
K.~Gutzmann, Access control and session management in the http environment,
  IEEE Internet Computing 5~(1) (2001) 26--35.

\bibitem{A59}
N.~Poggi, T.~Moreno, J.~L. Berral, R.~Gavalda, J.~Torres, Self-adaptive
  utility-based web session management, Computer Networks 53~(10) (2009)
  1712--1721.

\bibitem{A63}
K.~Czajkowski, I.~Foster, N.~Karonis, C.~Kesselman, S.~Martin, W.~Smith,
  S.~Tuecke, A resource management architecture for metacomputing systems, in:
  Workshop on Job Scheduling Strategies for Parallel Processing, Springer,
  1998, pp. 62--82.

\bibitem{A64}
H.~Kim, J.~Park, Y.~Jeong, Efficient resource management scheme for storage
  processing in cloud infrastructure with internet of things, Wireless Personal
  Communications 91~(4) (2016) 1635--1651.

\bibitem{A65}
D.~Zhang, Y.~Qiao, L.~She, R.~Shen, J.~Ren, Y.~Zhang, Two time-scale resource
  management for green internet of things networks, IEEE Internet of Things
  Journal 6~(1) (2018) 545--556.

\bibitem{A66}
H.~Yang, W.~Zhong, C.~Chen, A.~Alphones, X.~Xie,
  Deep-reinforcement-learning-based energy-efficient resource management for
  social and cognitive internet of things, IEEE Internet of Things Journal
  7~(6) (2020) 5677--5689.

\bibitem{A30}
J.~Lanier, A.~Heilbrun, A portrait of the young visionary, Whole Earth Review
  (1989) 108--19.

\bibitem{A29}
G.~C. Burdea, P.~Coiffet, Virtual reality technology, John Wiley \& Sons, 2003.

\bibitem{A32}
D.~Edler, O.~K{\"u}hne, J.~Keil, F.~Dickmann, Audiovisual cartography:
  Established and new multimedia approaches to represent soundscapes,
  KN-Journal of Cartography and Geographic Information 69~(1) (2019) 5--17.

\bibitem{A34}
M.~Chang, J.~Lambert, P.~Sangkloy, J.~Singh, , et~al., Argoverse: 3d tracking
  and forecasting with rich maps, in: Proceedings of the IEEE/CVF Conference on
  Computer Vision and Pattern Recognition, 2019, pp. 8748--8757.

\bibitem{A35}
T.~Li, W.~Wang, Using wearable devices to participate in 3d interactive
  storytelling, in: International Conference on Interactive Digital
  Storytelling, Springer, 2021, pp. 80--93.

\bibitem{A33}
W.~Mphepo, Stereoscopy and Autostereoscopy, Mixed Reality [Working Title],
  2020.

\bibitem{A31}
R.~Silva, J.~C. Oliveira, G.~A. Giraldi, Introduction to augmented reality,
  National laboratory for scientific computation 11 (2003) 1--11.

\bibitem{A36}
A.~Iatsyshyn, K.~Valeriia, Y.~O. Romanenko, I.~I. Deinega, S.~H. Lytvynova,
  Application of augmented reality technologies for preparation of specialists
  of new technological era, in: 2nd International Workshop on Augmented Reality
  in Education, AREdu 2019, 2019.

\bibitem{A37}
D.~Tome, P.~Peluse, L.~Agapito, H.~Badino, xr-egopose: Egocentric 3d human pose
  from an hmd camera, in: Proceedings of the IEEE/CVF International Conference
  on Computer Vision, 2019, pp. 7728--7738.

\bibitem{A38}
T.~Jung, M.~C. Tom~Dieck, Augmented reality and virtual reality, Ujedinjeno
  Kraljevstvo: Springer International Publishing AG.

\bibitem{A39}
S.~Rokhsaritalemi, A.~Sadeghiniaraki, S.~Choi, A review on mixed reality:
  Current trends, challenges and prospects, Applied Sciences 10~(2) (2020) 636.

\bibitem{A40}
A.~Rezeika, M.~Benda, P.~Stawicki, F.~Gembler, A.~Saboor, I.~Volosyak,
  Brain--computer interface spellers: A review, Brain sciences 8~(4) (2018) 57.

\bibitem{A41}
H.~J. Baek, M.~H. Chang, J.~Heo, K.~S. Park, Enhancing the usability of
  brain-computer interface systems, Computational intelligence and neuroscience
  2019.

\bibitem{A43}
R.~Winkler, Elon musk launches neuralink to connect brains with computers, The
  Wall Street Journal 27.

\bibitem{A42}
D.~Marshall, D.~Coyle, S.~Wilson, M.~Callaghan, Games, gameplay, and bci: the
  state of the art, IEEE Transactions on Computational Intelligence and AI in
  Games 5~(2) (2013) 82--99.

\bibitem{A44}
J.~Gregory, Game engine architecture, AK Peters/CRC Press, 2018.

\bibitem{A45}
R.~Salama, M.~ElSayed, Basic elements and characteristics of game engine,
  Global Journal of Computer Sciences: Theory and Research 8~(3) (2018)
  126--131.

\bibitem{A46}
C.~Fink, A new 3d game engine that means business, Available online:
  \url{https://www.forbes.com/sites/charliefink/2020/05/28/a-new-3d-game-engine-that-means-business/?sh=68bd4b9a7530}
  (accessed on 30 January 2022).

\bibitem{A60}
D.~A.~L. Carvajal, M.~M. Morita, G.~M. Bilmes, Virtual museums. captured
  reality and 3d modeling, Journal of Cultural Heritage 45 (2020) 234--239.

\bibitem{A61}
R.~Winzenrieth, L.~Humbert, S.~Di~Gregorio, E.~Bonel, M.~Garc{\'\i}a,
  L.~Del~Rio, Effects of osteoporosis drug treatments on cortical and
  trabecular bone in the femur using dxa-based 3d modeling, Osteoporosis
  International 29~(10) (2018) 2323--2333.

\bibitem{A62}
M.~Burgos, E.~Sanmiguel-Rojas, N.~Singh, F.~Esteban-Ortega,
  Digbody{\textregistered}: a new 3d modeling tool for nasal virtual surgery,
  Computers in biology and medicine 98 (2018) 118--125.

\bibitem{A73}
T.~Akeninemoller, E.~Haines, N.~Hoffman, Real-time rendering, AK Peters/crc
  Press, 2019.

\bibitem{A74}
G.~Wong, J.~Wang, Real-Time Rendering, CRC Press, 2013.

\bibitem{A75}
V.~Sanzharov, V.~A. Frolov, V.~A. Galaktionov, Survey of nvidia rtx technology,
  Programming and Computer Software 46~(4) (2020) 297--304.

\bibitem{A67}
S.~Mystakidis, Metaverse, Encyclopedia 2~(1) (2022) 486--497.

\bibitem{A68}
A.~T. Maereg, A.~Nagar, D.~Reid, E.~L. Secco, Wearable vibrotactile haptic
  device for stiffness discrimination during virtual interactions, Frontiers in
  Robotics and AI 4 (2017) 42.

\bibitem{A69}
X.~Jin, Y.~Yang, F.~JIANG, J.~Shuyuan, Social network structure feature
  analysis and its modelling [j], Bulletin of Chinese academy of sciences
  30~(2) (2015) 216--228.

\bibitem{A70}
M.~A. Smith, B.~Shneiderman, N.~Milic-Frayling, E.~Mendes~Rodrigues, V.~Barash,
  C.~Dunne, T.~Capone, A.~Perer, E.~Gleave, Analyzing (social media) networks
  with nodexl, in: Proceedings of the fourth international conference on
  Communities and technologies, 2009, pp. 255--264.

\bibitem{A71}
D.~J. Watts, S.~H. Strogatz, Collective dynamics of ‘small-world’networks,
  nature 393~(6684) (1998) 440--442.

\bibitem{A72}
A.~Barab{\'a}si, R.~Albert, Emergence of scaling in random networks, science
  286~(5439) (1999) 509--512.

\bibitem{A97}
A.~Mislove, M.~Marcon, K.~P. Gummadi, P.~Druschel, B.~Bhattacharjee,
  Measurement and analysis of online social networks, in: Proceedings of the
  7th ACM SIGCOMM conference on Internet measurement, 2007, pp. 29--42.

\bibitem{A98}
W.~Fan, Y.~Ma, Q.~Li, Y.~He, E.~Zhao, J.~Tang, D.~Yin, Graph neural networks
  for social recommendation, in: The world wide web conference, 2019, pp.
  417--426.

\bibitem{A99}
L.~Luceri, T.~Braun, S.~Giordano, Social influence (deep) learning for human
  behavior prediction, in: International Workshop on Complex Networks,
  Springer, 2018, pp. 261--269.

\bibitem{A100}
J.~Qiu, J.~Tang, H.~Ma, Y.~Dong, K.~Wang, J.~Tang, Deepinf: Social influence
  prediction with deep learning, in: Proceedings of the 24th ACM SIGKDD
  International Conference on Knowledge Discovery \& Data Mining, 2018, pp.
  2110--2119.

\bibitem{A76}
Z.~Yu, X.~Zhou, Socially Aware Computing: Concepts, Technologies, and
  Practices, Springer New York, New York, NY, 2014, pp. 9--23.

\bibitem{A79}
B.~Hull, V.~Bychkovsky, Y.~Zhang, K.~Chen, M.~Goraczko, A.~Miu, E.~Shih,
  H.~Balakrishnan, S.~Madden, Cartel: a distributed mobile sensor computing
  system, in: Proceedings of the 4th international conference on Embedded
  networked sensor systems, 2006, pp. 125--138.

\bibitem{A77}
Y.~Liu, K.~Wang, H.~Guo, Q.~Lu, Y.~Sun, Social-aware computing based congestion
  control in delay tolerant networks, Mobile Networks and Applications 22~(2)
  (2017) 174--185.

\bibitem{A78}
K.~Zhang, J.~Cao, H.~Liu, S.~Maharjan, Y.~Zhang, Deep reinforcement learning
  for social-aware edge computing and caching in urban informatics, IEEE
  Transactions on Industrial Informatics 16~(8) (2020) 5467--5477.
\newblock \href {http://dx.doi.org/10.1109/TII.2019.2953189}
  {\path{doi:10.1109/TII.2019.2953189}}.

\bibitem{A80}
N.~M, An overview of cognitive computing, International Journal of Swarm
  Intelligence and Evolutionary Computation 10~(3).

\bibitem{A81}
F.~F. Petiwala, V.~K. Shukla, S.~Vyas, Ibm watson: Redefining artificial
  intelligence through cognitive computing, in: Proceedings of International
  Conference on Machine Intelligence and Data Science Applications, Springer,
  2021, pp. 173--185.

\bibitem{A83}
A.~Sharma, A.~Sharma, J.~K. Pandey, M.~Ram, Swarm Intelligence: Foundation,
  Principles, and Engineering Applications, CRC Press, 2022.

\bibitem{A82}
G.~Beni, J.~Wang, Swarm intelligence in cellular robotic systems, in: Robots
  and biological systems: towards a new bionics?, Springer, 1993, pp. 703--712.

\bibitem{A85}
S.~Luo, L.~Cheng, B.~Ren, Practical swarm optimization based fault-tolerance
  algorithm for the internet of things, KSII Transactions on Internet and
  Information Systems (TIIS) 8~(3) (2014) 735--748.

\bibitem{A87}
O.~Zedadra, A.~Guerrieri, N.~Jouandeau, G.~Spezzano, H.~Seridi, G.~Fortino,
  Swarm intelligence-based algorithms within iot-based systems: A review,
  Journal of Parallel and Distributed Computing 122 (2018) 173--187.

\bibitem{A88}
J.~Garc{\'\i}anieto, E.~Alba, A.~C. Olivera, Swarm intelligence for traffic
  light scheduling: Application to real urban areas, Engineering Applications
  of Artificial Intelligence 25~(2) (2012) 274--283.

\bibitem{A86}
U.~Baum{\"o}l, R.~Jung, B.~J. Kr{\"a}mer, Advances in collective intelligence
  and social media (2013).

\bibitem{A84}
D.~Chen, R.~Zhang, Exploring research trends of emerging technologies in health
  metaverse: A bibliometric analysis, Available at SSRN 3998068.

\end{thebibliography}
~~~\\
~~~\\




 


\end{document}